%% file: Article.tex
\documentclass[usenatbib]{mn2e}
\pdfoutput=1

\usepackage{times}

\usepackage{graphicx}

\usepackage{amssymb}
\usepackage{amsmath}

\usepackage{color}      
\definecolor{dark-grey}{gray}{0.5}

\usepackage[
  colorlinks=true,     
  citecolor=dark-grey, 
  linkcolor=dark-grey  
]{hyperref}
\hypersetup{
pdfauthor={Sander Valcke}, 
pdftitle={Valcke et al. Kelvin-Helmholtz instabilities in Smoothed
  Particle Hydrodynamics}
}

\graphicspath{{images/highres/}{images/elke/}}

\newcommand{\rs}[1]{\ensuremath{_\mathrm{#1}}}
\newcommand{\difrac}[2]{\frac{\mathrm{d}#1}{\mathrm{d}#2}}

\newcommand{\pfrac}[2]{\frac{{\partial}{#1}}{\partial{#2}}}
\newcommand{\ppfrac}[2]{\frac{{\partial}^2{#1}}{\partial{#2}^2}}
\newcommand{\atan}{\ensuremath{\mathrm{atan}}}
\newcommand{\vt}[1]{\ensuremath{\bmath{#1}}}

\title[Kelvin-Helmholtz instabilities in SPH]{Kelvin-Helmholtz instabilities in Smoothed Particle Hydrodynamics}

\author[S. Valcke et al.]{S. Valcke$^{1}$\thanks{Doctoral Fellow of
    the Fund for Scientific Research -- Flanders, Belgium
    (FWO). E-mail: Sander.Valcke@UGent.be}, S. De
  Rijcke$^{1}$\thanks{E-mail: Sven.Derijcke@UGent.be},
  E. R\"odiger$^2$\& H. Dejonghe$^1$\\ $^{1}$Sterrenkundig
  Observatorium, Ghent University, Krijgslaan 281, S9, 9000 Gent,
  Belgium\\ $^2$Jacobs University Bremen, PO Box 750 561, 28725 Bremen, Germany}
\begin{document}
\date{Accepted . Received ; in original form }
\pagerange{\pageref{firstpage}--\pageref{lastpage}} \pubyear{2009}
\maketitle
\label{firstpage}
\begin{abstract}
\input{abstract}
\end{abstract}
\begin{keywords}
hydrodynamics -- instabilities -- methods: numerical
keywords
\end{keywords}
\input{intro}
\input{theCode}
\input{liqKernel}
\input{shearingLayers}
\input{discussion}
\section*{Acknowledgments}
We thank the referee for the valuable and constructive comments. We
also thank Justin Read for making the code to compute the $E_0$ errors
publicly
available\footnote{http://www.astrosim.net/code/doku.php?id=codesonline:nbody:osphpatch}. SV
acknowledges and is grateful for the financial support of the Fund for
Scientific Research -- Flanders (FWO). The SPH simulations were run on
our local computer cluster ITHILDIN. The \textsc{FLASH} software used
for one of the simulations in this paper was developed by the DOE
supported ASCI/Alliances Center for Astrophysical Thermonuclear
Flashes at the University of Chicago.

\bibliographystyle{mn2e} 
\bibliography{biblio}
\appendix
\input{appendix}
\label{lastpage}
\end{document}

%% file: abstract.tex
In this paper we investigate whether Smoothed Particle Hydrodynamics
(SPH), equipped with artificial conductivity, is able to capture the
physics of density/energy discontinuities in the case of the so-called
shearing layers test, a test for examining Kelvin-Helmholtz (KH)
instabilities.  We can trace back each failure of SPH to show KH rolls
to two causes: i) shock waves travelling in the simulation box and ii)
particle clumping, or more generally, particle noise. The probable cause
of shock waves is the Local Mixing Instability (LMI), previously
identified in the literature. Particle noise on the other hand is a
problem because it introduces a large error in the SPH momentum
equation.

The shocks are hard to avoid in SPH simulations with initial density
gradients because the most straightforward way of removing them, i.e.
relaxing the initial conditions, is not viable. Indeed, by the time
sufficient relaxing has taken place the density and energy gradients
have become prohibitively wide. The particle disorder introduced by
the relaxation is also a problem. We show that setting up initial
conditions with a suitably smoothed density gradient dramatically
improves results: shock waves are reduced whilst retaining relatively
sharp gradients and avoiding unnecessary particle disorder. Particle
clumping is easy to overcome, the most straightforward method being
the use of a suitable smoothing kernel with non-zero first central
derivative. We present results to that effect using a new smoothing
kernel: the LInear Quartic (LIQ) kernel.

We also investigate the role of artificial conductivity (AC). Although
AC is necessary in the simulations to avoid ``oily'' features in the
gas due to artificial surface tension, we fail to find any relation
between using artificial conductivity and the appearance of seeded KH
rolls. Including AC is necessary for the long-term behavior of the
simulation (e.g. to get $\lambda=1/2, 1$ KH rolls). In sensitive
hydrodynamical simulations great care is however needed in selecting
the AC signal velocity, with the default formulation leading to too
much energy diffusion. We present new signal velocities that lead to
less diffusion.

The effects of the shock waves and of particle disorder become less
important as the time-scale of the physical problem (for the shearing
layers problem: lower density contrast and higher Mach numbers)
decreases. At the resolution of current galaxy formation simulations
mixing is probably not important. However, mixing could become crucial
for next-generation simulations.

%% file: intro.tex
\section{Introduction}
Smoothed Particle Hydrodynamics (SPH) is a lagrangian technique to
solve the equations of hydrodynamics. It was first conceived by
\citet{lucy1977} and \citet{gingold1977} to solve astrophysical
problems. Because the differential equations are solved on a particle
mesh that moves with the flow, adaptivity to accomodate large density
variations is inherent to SPH. As shown in \citet{tasker2008} SPH
indeed has the advantage over grid-based techniques when it comes to
problems with a large dynamic range. However, when it comes to
handling steep density gradients SPH codes have to acknowledge their
superiors in grid-based codes.

Recently, doubt has been cast upon whether SPH is able to capture all of
the physics related to discontinuities. \citet{agertz2007} highlighted a
known problem in SPH: the formation of an artificial gap around density
discontinuities. Previous allusions to this problem are found in
e.g. \citet{cummins1999}, where a method to enforce incompressibility in
SPH is proposed. \citet{tartakovsky2005} propose a method to overcome
this problem: use the local number density instead of the normal density
to weigh SPH variables. This shifts the problem from requiring a
continuous density to a continuous number density. The latter can be
achieved by setting up the initial conditions such that the low density
fluid uses SPH particles with lower mass, instead of less SPH particles
with the same mass. \citet{agertz2007} furthermore argue that a basic
SPH scheme is unable to exhibit Kelvin-Helmholtz or Rayleigh-Taylor
instabilities when a density gradient is involved. In a follow-up paper,
\citet{read2010} identify two main problems with standard SPH
implementations: the Local Mixing Instability (LMI) and the ``E0'' error
in the momentum equation. The LMI is the cause of the artificial gap
problem highlighted above, whereas the problem with the E0 error in the
SPH momentum equation has previously been highlighted by
\citet{morris1996}. \citet{read2010} present a solution to these
problems based on a temperature-weighted density and a modified
smoothing kernel.

\citet{price2008} presents a different solution to the artificial gap
problem. Based on the general approach previously outlined by
\citet{monaghan1997} he introduces an ``artificial conductivity'' (AC)
term into the equations. This term induces a certain amount of energy
diffusion across energy discontinuities, allowing a discontinuity (which
SPH is unable to handle properly) to become more ``smeared out'' and
thus treatable with SPH.

\citet{kawata2009} present their implementation of an SPH scheme,
closely tailored after the scheme by \citet{rosswog2007}, which includes
the artificial conductivity term. They present the basic tests performed
by \citet{agertz2007}: the blob test and the shearing layers test. Their
results indicate that the \citet{price2008} AC solution gives improved
performance for the blob test, on the condition that enough particles
are present in the blob. Their shearing layers test (with a density
contrast of 9.6) exhibits Kelvin-Helmoltz instabilities for a
Kelvin-Helmholtz time-scale of $\tau\rs{KH}=0.57$.

Interestingly, \citet{okamoto2003} investigated shearing flows in
SPH. They find that noisiness in the SPH smoothing of variables gives
rise to small-scale pressure gradients which significantly decelerate
the shearing flow.

In \S~\ref{sec:theCode} we give a brief introduction to the SPH code,
highlighting the implementation of artificial conductivity (AC). A new
smoothing kernel with non-zero central first derivative is presented
in \S~\ref{sec:liqKernel}. Various aspects of the shearing layers
test are then examined in \S~\ref{sec:shearingLayers} and a discussion
is given in \S~\ref{sec:discussion}.

Most of the plots in this paper were made using HYPLOT. HYPLOT is a
freely
available\footnote{http://sourceforge.net/apps/wordpress/hyplot/about/}
open source analysis package, with an emphasis on SPH. Currently only
\textsc{GADGETII} file reads are supported. HYPLOT uses PyQt4 for its
GUI front\-end, matplotlib for the plotting and a host of C++ classes
for the actual computations. Interested users are recommended to get the
latest snapshot from the svn
repository\footnote{https://hyplot.svn.sourceforge.net/svnroot/hyplot/trunk}.

%% file: theCode.tex
\section{The Code}
\label{sec:theCode}
We use the publicly available version of the Nbody/SPH code GADGETII
\citep{springel2005}. There are two base premises when formulating a
Smoothed Particle Hydrodynamics (SPH) solution to the equations of
hydrodynamics:
\begin{enumerate}
  \item the integral representation of field functions,
  \item the particle approximation.
\end{enumerate}
In the first premise a function is replaced by its integral
representation, given by the integration of the multiplication of that
function and a smoothing kernel function. The second premise states
that the integral from the first premise is replaced by a discretized
summation using a set of particles in the support domain. The latter
is a key approximation as it obsoletes the use of a background mesh
for numerical integration. Both premises allow us to use the following
simple equation to calculate the density at a certain point in space
$\vt{r}$:
\begin{equation}
  \rho(\vt{r}) = \sum_{i=1}^N m_{j} W(|\vt{r}-\vt{r}_i|, h),
\end{equation}
where the summation goes over all the particles within the support
domain, delimited by the smoothing length $h$. Here, $m_{j}$ is the
mass of the $j$th particle, $W$ is a smoothing function.

For a derivation of the SPH formulation of the basic equations of
hydrodynamics we refer the reader to e.g. \citet{springel2002}, where
the equations are derived from a Lagrangian variational
principle. 
\subsection{Artificial Conductivity}
\label{subsec:theCode_AC}
From e.g. \citet{monaghan1997} and \citet{price2008} we learn that an
artificial conductivity (AC) term should be included in the SPH
equations when dealing with energy discontinuities. The expression for
the dissipational part (i.e. without the adiabatic part) of the energy
equation for an SPH particle $i$ then becomes:
\begin{align}
\left( \difrac{u_{i}}{t} \right) \rs{diss} = &
 \sum_j \frac{m_{j}}{\bar{\rho}_{ij}} \Big[ - \frac{1}{2}\alpha v\rs{sig}(\vt{v}_{ij}\cdot\hat{\vt{e}}_{ij})^2 \nonumber \\
   & \qquad + \alpha_u v\rs{sig}^u(u_{i} - u_{j}) \Big]
 \hat{\vt{e}}_{ij} \cdot \overline{\vt{\nabla}_{i}W_{ij}},
\label{eq:dudtdiss}
\end{align}
with $u$ the specific energy. The summation over $j$ is the sum over
all neighbours of particle $i$. Here, $m_j$ is the neighbour particle
mass, $\bar{\rho}_{ij}=(\rho_i+\rho_j)/2$ with $\rho_i$ the standard
SPH particle density. $\alpha$ and $\alpha_u$ are coefficients that
can be used to dynamically vary the contribution of the terms based on
the presence of e.g. local velocity convergence. We set $\alpha =
\alpha_u = 1$ and
$v\rs{sig}=(c_i+c_j-\beta\vt{v}_{ij}\cdot\hat{\vt{e}}_{ij})$, with
$c_i$ the sound speed of particle $i$. Note that the first term in
equation (\ref{eq:dudtdiss}) with $\beta$ set to $1.5$ is equal to the
artificial viscosity (AV) employed in the GADGETII code
\citep{springel2005}, apart from the Balsara switch. The second term
in eq. (\ref{eq:dudtdiss}) is the artificial conductivity (AC) term.

\citet{price2008} suggests the following form
for the AC signal velocity $v\rs{sig}^u$:
\begin{equation}
  v\rs{sig}^u = \sqrt{ \frac{|P_{i} - P_{j}|}{\bar{\rho}_{ij}}},
  \label{eq:pricevsig}
\end{equation}
with $P_{i}$ and $P_{j}$ the pressures of respectively particles $i$
and $j$. With this choice, spurious pressure gradients across contact
discontinuities are gradually eliminated. We note however that as the
expression for the artificial conductivity (\ref{eq:dudtdiss}) is
actually an SPH representation of a diffusion term \citep{price2008},
using the suggested signal velocity (eq. (\ref{eq:pricevsig})) could
lead to spurious energy diffusion. When using this signal velocity to
eliminate pressure discontinuities one implicitly assumes that lower
energy corresponds with lower pressure, as the sign of the energy
transfer is determined by the energy gradient whilst the magnitude of
the transfer is determined by both the energy and pressure
gradients. If this assumption is valid the diffusion will distribute
energy from the high-energy particles to the low-energy particles,
reducing the pressure discontinuities. In the simulations we use the
equation of state for an ideal gas:
\begin{equation}
  p = (\gamma - 1) \rho u,
  \label{eq:idgaseqstate}
\end{equation}
with $p$ the pressure, $\gamma$ the adiabatic constant (typically
taken to be 5/3, the value for an ideal mono-atomic gas). From
eq. (\ref{eq:idgaseqstate}) we learn that we can only know that higher
energy corresponds to higher pressure when the density is constant. If
the {\em low-energy particles have higher pressure}, the situation
will not reach the desired pressure equilibrium as energy will flow
from the high-energy particles to the low-energy particles, {\em
  increasing the pressure discontinuities}. In this case energy will
keep flowing until an approximate energy equilibrium is reached. It is
straightforward to formulate a signal velocity that does not suffer
from this problem:
\begin{equation}
  v\rs{sig,1}^u = \mathrm{sign}\left[(P_{i} - P_{j})(u_{i} - u_{j})\right]\sqrt{ \frac{|P_{i} - P_{j}|}{\bar{\rho}_{ij}}},
  \label{eq:ownvsig}
\end{equation}
where the sign of the energy diffusion is now determined by the sign
of the pressure gradient. Expanding on this idea, another possible
form of the signal velocity is:
\begin{equation}
  v\rs{sig,2}^u = \mathrm{sign}\left[(\tilde{P}_{i} - \tilde{P}_{j})(u_{i} - u_{j})\right]\sqrt{ \frac{|\tilde{P}_{i} - \tilde{P}_{j}|}{\bar{\rho}_{ij}}},
  \label{eq:ownvsig2}
\end{equation}
where $\tilde{P}$ is the SPH-averaged pressure:
\begin{equation}
  \tilde{P}_i = \sum_{i=1}^N m\rs{j} A_j \rho^{\gamma-1}_j \bar{W}(|\vt{r}-\vt{r}_i|, h),
  \label{eq:sump}
\end{equation}
with $A_j$ the entropy of particle $j$, $\bar{W} = (W_i+W_j)/2$. The
advantage of this form is that the artificial conductivity counteracts
pressure differences at the fluid level, not at the particle level. A
disadvantage is the extra overhead needed to compute and store
$\tilde{P}$. Note that (as highlighted by \citet{price2008}) the
formulations for the signal velocity shown here
(eqs. (\ref{eq:pricevsig}), (\ref{eq:ownvsig}) and (\ref{eq:ownvsig2}))
are not applicable when gravity is involved, because typically under
hydrostatic equilibrium a configuration with constant pressure is not an
equilibrium configuration. A signal velocity applicable in an
environment with gravity, e.g. a modified version of
eq. (\ref{eq:pricevsig}), can also be extended with the factor we added
here in going from eq. (\ref{eq:pricevsig}) to eq. (\ref{eq:ownvsig}).

\begin{figure}
\includegraphics[width=\hsize]{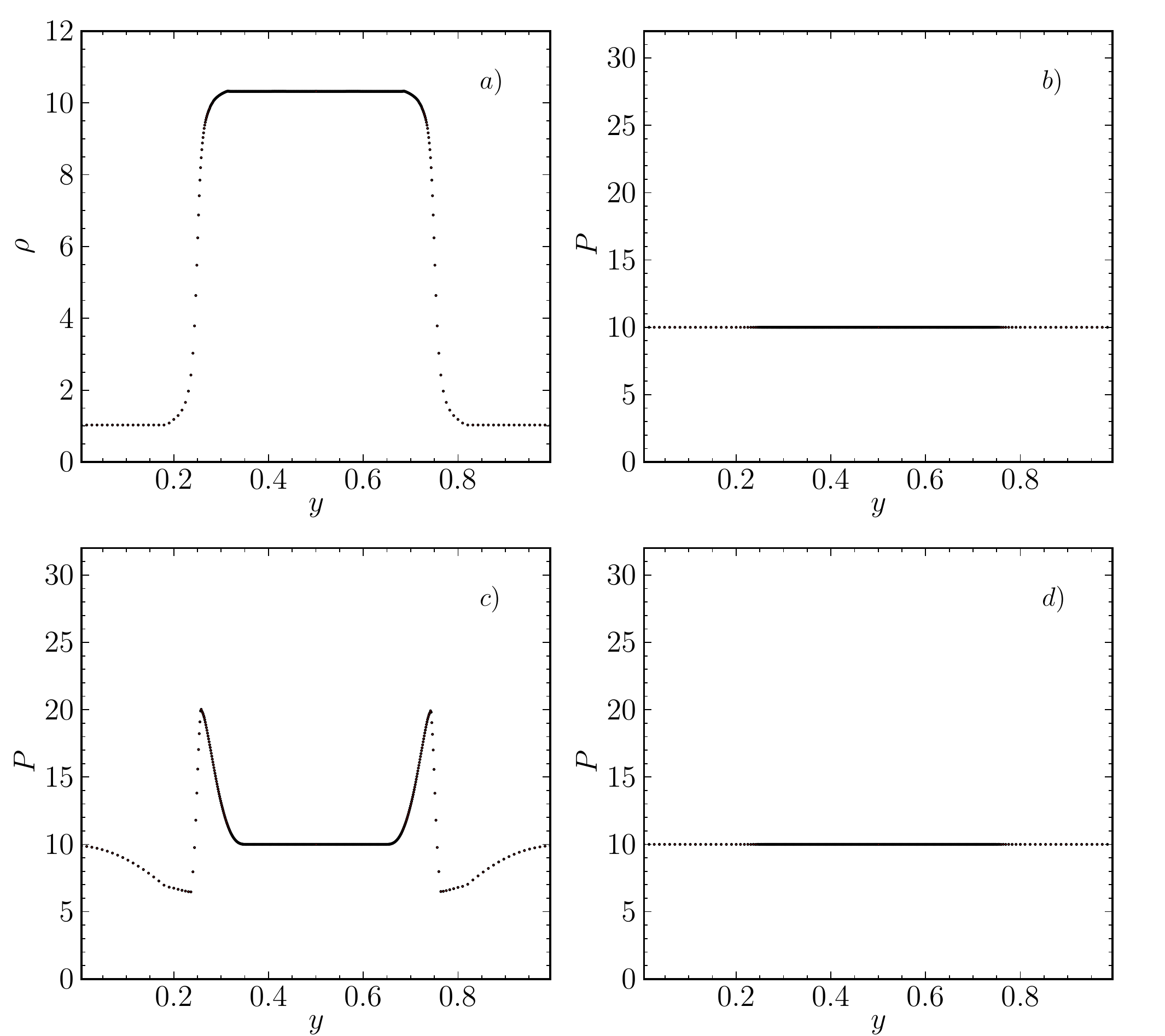}
\caption{{\em a)} Initial density along the $y$-axis for an $x$-slice
  [0,0.001]. {\em b)} Initial pressure along the $y$-axis. {\em c)}
  Pressure along the $y$-axis at time $t=2$ in the case of the
  standard AC signal velocity (eq. (\ref{eq:pricevsig})). {\em d)}
  Pressure along the $y$-axis at time $t=2$ in the case of the
  modified AC signal velocity (eq. (\ref{eq:ownvsig})).
\label{fig:fixed_particles_svenzograms}}
\end{figure}

It is worth considering whether the potential errors induced by
equation (\ref{eq:pricevsig}) actually happen, and if so if their
magnitude is of an order that requires fixing. To test this we set up
two simulations, one using the signal speed of
eq. (\ref{eq:pricevsig}), the other using
eq. (\ref{eq:ownvsig}). These simulations consist of two central
smoothed density gradients and constant pressure, exactly the same way
we set up simulations for the shearing layers tests (section
\ref{sec:shearingLayers}). Initial specific energies are set on a
per-particle basis using the calculated SPH density and a fixed
constant pressure value. To be able to clearly examine the influence
of the conductivity the SPH particles are fixed in place. Results are
shown in Fig. \ref{fig:fixed_particles_svenzograms}.
We immediately see that the simulation with the signal velocity as in
eq. (\ref{eq:pricevsig}) (panel $c$) indeed exhibits divergent
behavior. SPH particles with low energy ($0.25 < x < 0.75$) receive
extra energy, increasing their pressure, and vice versa for particles
with high energy. Once this mechanism is set in motion it is
self-amplifying. The simulation with the modified signal velocity on
the other hand (panel $d$) does not exhibit any special behavior, the
pressure remains constant, as expected. Note that the reason the
divergent behavior is set in motion in this setup is numerical
noise. Indeed, from eq. (\ref{eq:pricevsig}) one would expect a
simulation with equal particle pressures to have a signal velocity
which is equal to zero everywhere. As we set the initial particle
specific energies based on the required constant pressure value ($P =
10$) and the calculated density, we can expect that calculating back
to the pressure later on (again using eq. (\ref{eq:idgaseqstate}))
will yield non-zero albeit very small pressure differences. This does
not imply that the unwanted energy diffusion is a result of numerical
artifacts, because in a real simulation there will always be non-zero
pressure differences and we can expect at any time to have too much
energy diffusion. On the other hand, because in a real simulation the
SPH particles are not fixed in place they will react to the increased
pressure gradient, trying to erase it. The buildup of diffusion will
thus not be as drastical as the one shown in
Fig. \ref{fig:fixed_particles_svenzograms}. We will investigate the
new forms of the AC signal velocities further on.

%% file: liqKernel.tex
\section{LIQ Kernel}
\label{sec:liqKernel} A vital part of any SPH code is the smoothing
kernel $W$. The attention given to the smoothing kernel is however
somewhat limited, with people generally using the cubic spline kernel
\citep[see e.g.][]{kawata2009}. \citet{schuessler1981} already
demonstrated that the choice of the smoothing kernel can have a large
impact on simulations. More recently, \citet{morris1996, price2005,
read2010} performed a linear perturbation analysis of the SPH equations
of motion for different kernels, examining the stability of these
kernels for longitudinal (related to the clumping instability) and
transversal waves. They show several smoothing kernels with a zero
central derivative that suffer from the {\em clumping instability} (also
dubbed {\em tensile instability}).

Various approaches have been suggested to deal with this. One
possiblity is to modify the SPH equations. \citet{monaghan2000}
advocates including an artificial pressure term. \citet{sigalotti2009}
on the other hand apply an adaptive kernel density estimation
algorithm (ADKE). A different approach is to take a different form for
the smoothing kernel. The advantage of the latter approach is that no
modifications to the actual SPH scheme are necessary. The disadvantage
is that bell-shaped kernels (with central derivative equal to zero)
tend to give better results for a wide range of test cases as compared
to hyperbolic or parabolic shaped kernels \citep{fulk1997}.

Here we construct a new smoothing kernel with non-zero central
derivative: the LInear Quartic or LIQ kernel. We note several earlier
approaches: the HYDRA kernel by \citet{thomas1992}, who artificially
modified the cubic spline (CS) kernel to have a constant central first
derivative (by fixing it to $-1/\pi$ for $x \leq 2/3$). This approach
was recently picked up by \citet{merlin2009}. \citet{johnson1996}
employed a quadratic kernel, resulting in a linear first
derivative. More recently \citet{read2010} modified the central part
of the cubic spline kernel with their Core-Triangle (CT) kernel,
giving it a non-zero central first derivative. We will come back to
these at the end of section \ref{subsec:fixingxs}.

We assume a kernel of the form:
\begin{equation}
  W(\vt{r}, h) = \frac{W\rs{r}(u)}{h^d},
\end{equation}
with $u=r/h$ and $d$ the number of dimensions. For $W\rs{r}$
we take the following functional form:
\begin{equation}
  \label{eq:liqKernel}
  W\rs{r} = N \times
  \begin{cases}
    f_1: F - u & 0 \leq u < x\rs{s} \\
    f_2: A u^4 + B u^3 + C u^2 + D u + E & x\rs{s} \leq u < 1 \\
    0 & 1 \leq u
  \end{cases}.
\end{equation}
$x\rs{s}$ is a free parameter determining the connection point of the
polynomial and linear functions. This form is inspired by two ideas:
(i) we want the smoothing kernel to be smooth (ii) the first
derivative of the smoothing kernel should be a monotonously ascending
function (i.e. it can have constant parts, but it can never descend).

The equations for the smoothing kernel (\ref{eq:liqKernel}) have 7
free parameters: $A$ through $F$ and the normalization factor $N$. We
fix them by imposing the following boundary conditions:
\begin{subequations}
\label{eq:conditions}
\begin{eqnarray}
  f_2(x\rs{s}) & = & f_1(x\rs{s}) \label{eq:liq_firstEq}\\
  \pfrac{f_2}{u}(x\rs{s})  & = & \pfrac{f_1}{u}(x\rs{s}) \\
  \ppfrac{f_2}{u}(x\rs{s}) & = & 0 \\
  f_2(1)             & = & 0 \\
  \pfrac{f_2}{u}(1)  & = & 0 \\
  \ppfrac{f_2}{u}(1) & = & 0 \label{eq:liq_lastEq}\\
  \int_V  W(\vt{r}, h) \ \mathrm{d\vt{r}} & = & 1,
\end{eqnarray}
\end{subequations}
which ensure a smooth (up to second order) transition between $f_1$
and $f_2$ at $x\rs{s}$ and sufficiently smooth behavior for
$f_2\rightarrow 0$ as $u\rightarrow 1$.

\begin{figure}
\includegraphics[width=\hsize]{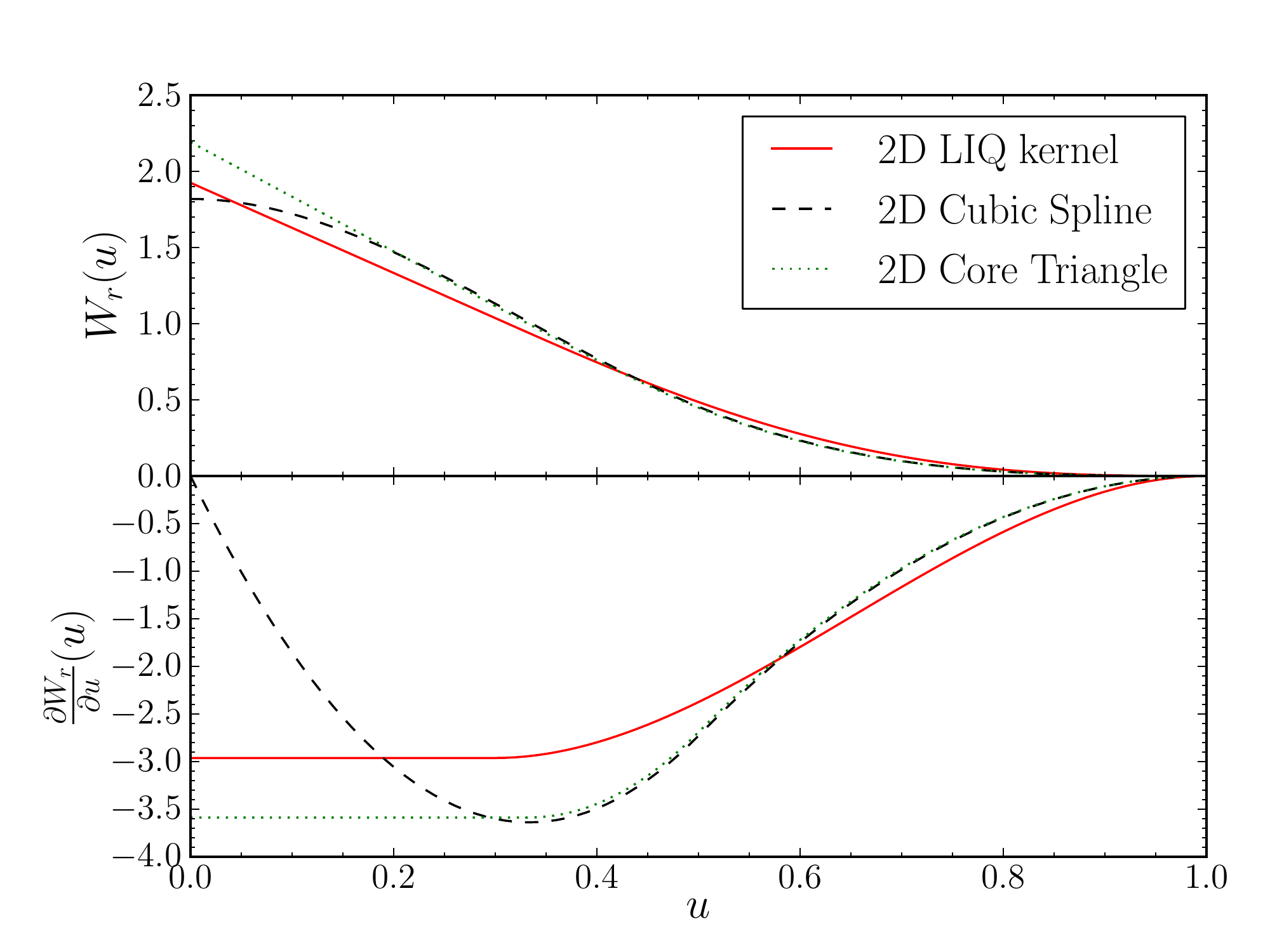}
\caption{{\em Upper panel}. The 2D LIQ, CS and CT smoothing
  kernels. {\em Lower panel}. First derivative of these
  kernels.  \label{fig:kernels}}
\end{figure}

\begin{figure}
\includegraphics[width=\hsize]{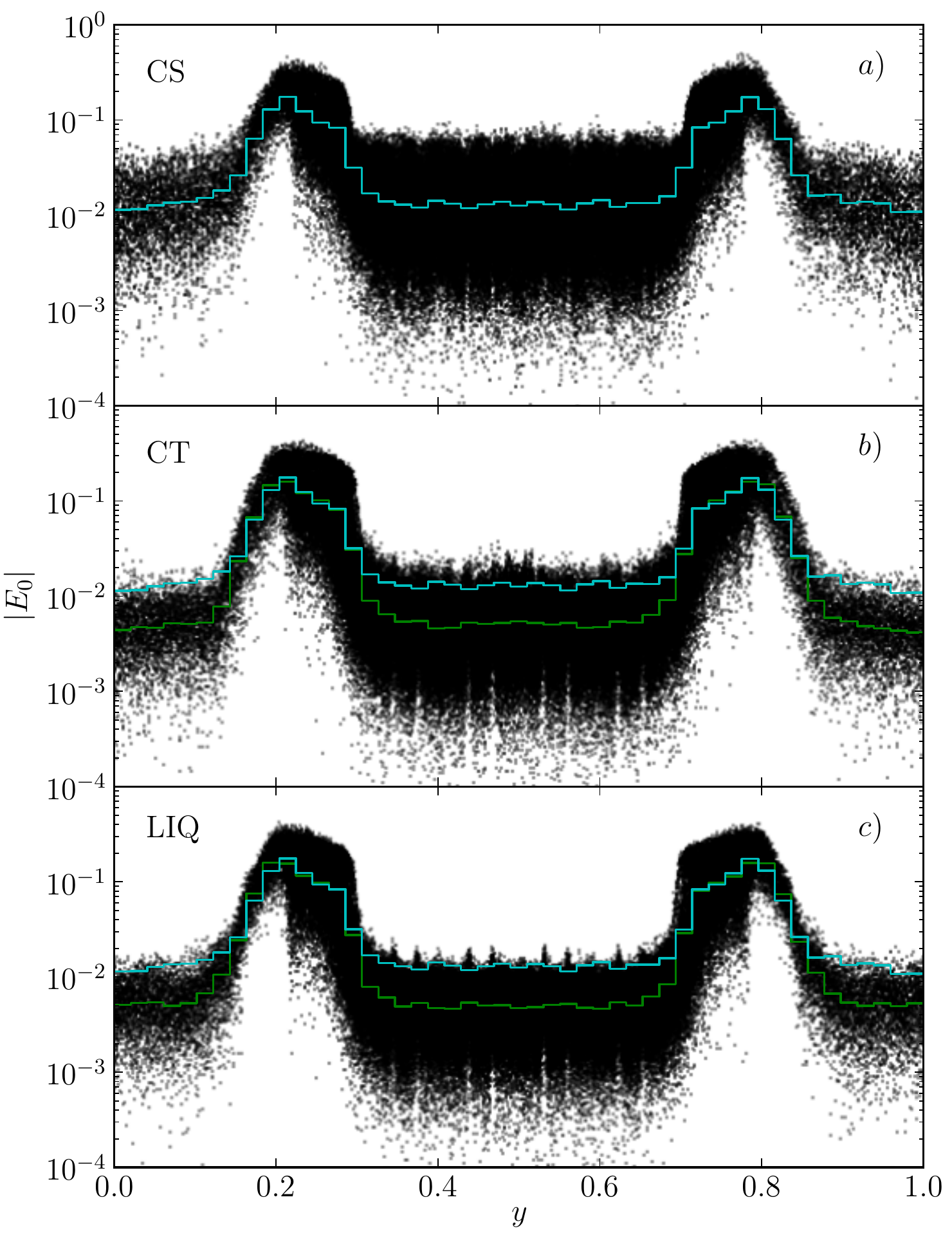}
\caption{The $E_0$ error for three shearing layers simulations ($t=0.5$)
with, from top to bottom: the cubic spline kernel, the core triangle
kernel, the linear quartic kernel. To guide the eye a binned mean value
is also shown. The cyan line, shown in all panels, shows the values for
the CS kernel. The green lines in panels $b$ and $c$ show the values for
the respective CT and LIQ kernels.\label{fig:E0error}}
\end{figure}

Three different kernels are shown in Fig. \ref{fig:kernels} in the
twodimensional case. The LIQ kernel has a lower central value than the
CT kernel, and a higher value outwards. \citet{read2010} find
  that they need a large number of neighbors for the CT kernel to
  reduce the $E_0$ error of the momentum equation and to reduce the
  pressure blips at the boundaries. The $E_0$ error on particle $i$ is
  given by \citep{read2010}:
\begin{equation}
  \vt{E}_{0,i} = \sum_j \frac{m_j}{\rho_j}\left(\frac{\rho_i}{\rho_j} + \frac{\rho_j}{\rho_i}\right) \overline{h \vt{\nabla} W}
\end{equation}
Fig. \ref{fig:E0error} shows a plot of particle $E_0$ errors as a
function of $y$, for the CS, CT and LIQ kernels. The number of neighbors
is kept equal (32), the setup is that of the RHO2 simulation (see Table
\ref{tab:khsims2} and Section \ref{sec:shearingLayers}). The difference
between the CT and LIQ kernels on the one hand and the CS kernel on the
other hand is big, with particle clumping giving rise to large $E_0$
errors, evidenced by the large scatter between the peaks. The results
for the CT and LIQ kernels are comparable, with the LIQ kernel slightly
in front, as can be seen from the reduced scatter away from the peak
areas and the smaller maximum values overall. It would be interesting to
perform the 3D simulations as performed by \citep{read2010} with the LIQ
kernel, in this paper we restrict ourselves to 2D however.  Both the LIQ
and CT kernels have the desired properties for their respective first
derivatives. Table \ref{tab:kernelCoeff} lists computed LIQ kernel
coefficients for a range of $x\rs{s}$ values. Analytical expressions for
the coefficients can be found in appendix \ref{sec:appendixLiq}.
\begin{table*}
\begin{minipage}{95mm}
\caption{Coefficients of the LIQ Kernel
  (eq. (\ref{eq:liqKernel})). The last two columns ($N\rs{2D}$,
  $N\rs{3D}$) give the norm $N$ respectively for two and three
  dimensions. See appendix \ref{sec:appendixLiq} for analytical
  expressions.\label{tab:kernelCoeff}}
\centering
\begin{tabular}{ccccccccc}
  $x\rs{s} $& $A$ & $B$ & $C$ & $D$ & $E$ & $F$ & $N\rs{2D}$ & $N\rs{3D}$\\
  \hline\hline
  0   & -0.5    & 1     &  0      & -1      & 05     & 0.5    & 4.775 & 6.685 \\
  0.2 & -0.9766 & 2.344 & -1.172  & -0.7813 & 0.5859 & 0.6    & 3.490 & 4.753 \\
  0.3 & -1.458  & 3.790 & -2.624  & -0.2915 & 0.5831 & 0.6500 & 2.962 & 3.947 \\
  0.4 & -2.315  & 6.481 & -5.556  &  0.9259 & 0.4630 & 0.7    & 2.508 & 3.251 \\
  0.6 & -7.813  & 25    & -28.125 &  12.5   & -1.563 & 0.8    & 1.798 & 2.168 \\
  0.8 & -62.5   & 225   & -300    & 175     & -37.5  & 0.9    & 1.300 & 1.434 \\
  \hline
\end{tabular}
\end{minipage}
\end{table*}
\subsection{Fixing $x\rs{s}$}
\label{subsec:fixingxs} 
\begin{figure}
\includegraphics[width=\hsize]{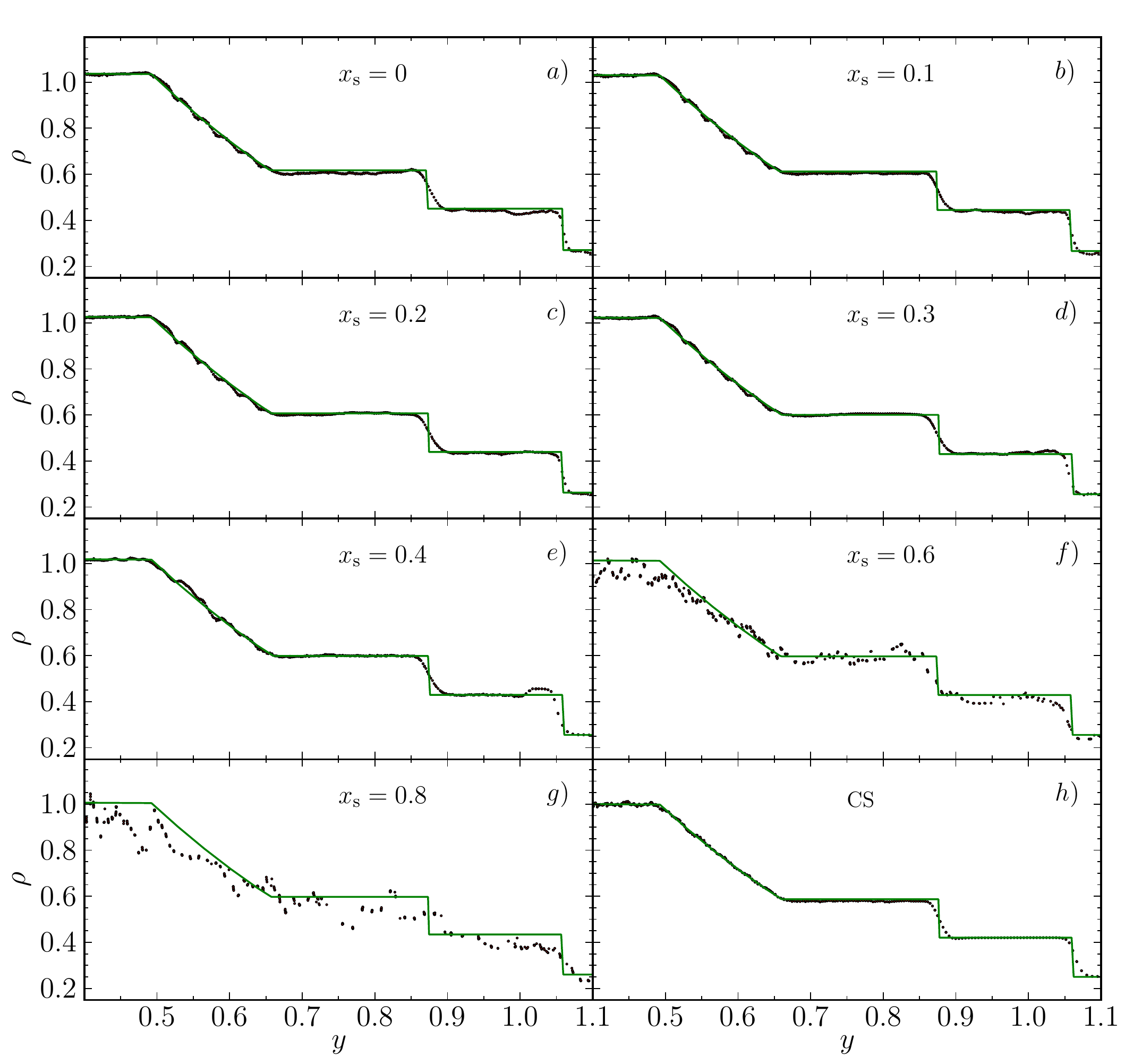}
  \caption{Results for the 2D Sod shock tube test with the LIQ Kernel,
    using varying connection points ($x\rs{s}$). Only particles in the
    $x$-interval [0,0.005] are shown ($x$-range: [0,0.1]). Connection
    point values are shown in the plots. Panel $h)$ shows the result
    using the standard Cubic Spline kernel. The overplotted solid
    green line in each plot is the analytical
    solution.\label{fig:SodShockTube}}
\end{figure}

To fix the remaining free parameter in the LIQ
Kernel, the connection point $x\rs{s}$, we set up a series of 2D Sod
shock tube tests (\citet{sod1978}, for recent use in SPH see
e.g. \citet{price2008} (1D)). The $x$-range is $[0-0.1]$, the $y$-range
$[0-1.5]$. The adiabatic index $\gamma$ is set to $5/3$. Further
parameters can be found in table \ref{tab:shocktube}. Throughout this
paper the number of SPH neighbours is 32, unless noted otherwise. Note
that this number is actually quite high, with about 5--6 neighbours per
dimension. The particles are set up on rows in the $y$-direction
i.e. the initial $x$-separation between particles is equal everywhere.
Results are shown in figure \ref{fig:SodShockTube}. From these results
we learn that varying $x\rs{s}$ has a major impact on the simulation. A
value of 0.3 is optimal, with higher values resulting in a more noisy
density profile and smaller values leading to small density oscillations
in certain parts of the profile. The shock tube result with the LIQ
kernel and $x\rs{s}=0.3$ is slightly less good than that for the CS
kernel (bottom right panel). The density found by the SPH summation
using the LIQ kernel is less accurate than the density found using the
CS kernel. The simulations in figure \ref{fig:SodShockTube} started from
the same initial conditions file, but have, depending on the value of
$x\rs{s}$, slightly different starting densities: the density is
overestimated, the more so for smaller values of $x\rs{s}$. Instead of
lowering the particle masses to get identical initial density profiles
we use identical masses in all simulations. The analytical shock-tube
results shown in Fig. \ref{fig:SodShockTube} are computed using the
actual initial SPH densities.
\begin{table}
  \begin{minipage}{68mm}
  \caption{Sod shock tube test parameters. (1) density (2) specific
    energy (3) pressure (4) number of particles per
    $y$-column (5) total number of particles \label{tab:shocktube}}
  \begin{tabular}{cccccc}
                 & $\rho$ & $u$  & $P$    & $N\rs{col}$ & $N$ \\
    \hline\hline
    high-density & 1      & 1.5  & 1      & 400         & 20000\\
    low density  & 0.25   & 1    & 0.1667 & 100         & 5000\\
    \hline
  \end{tabular}
  \end{minipage}
\end{table}

\begin{figure}
  \includegraphics[width=\hsize]{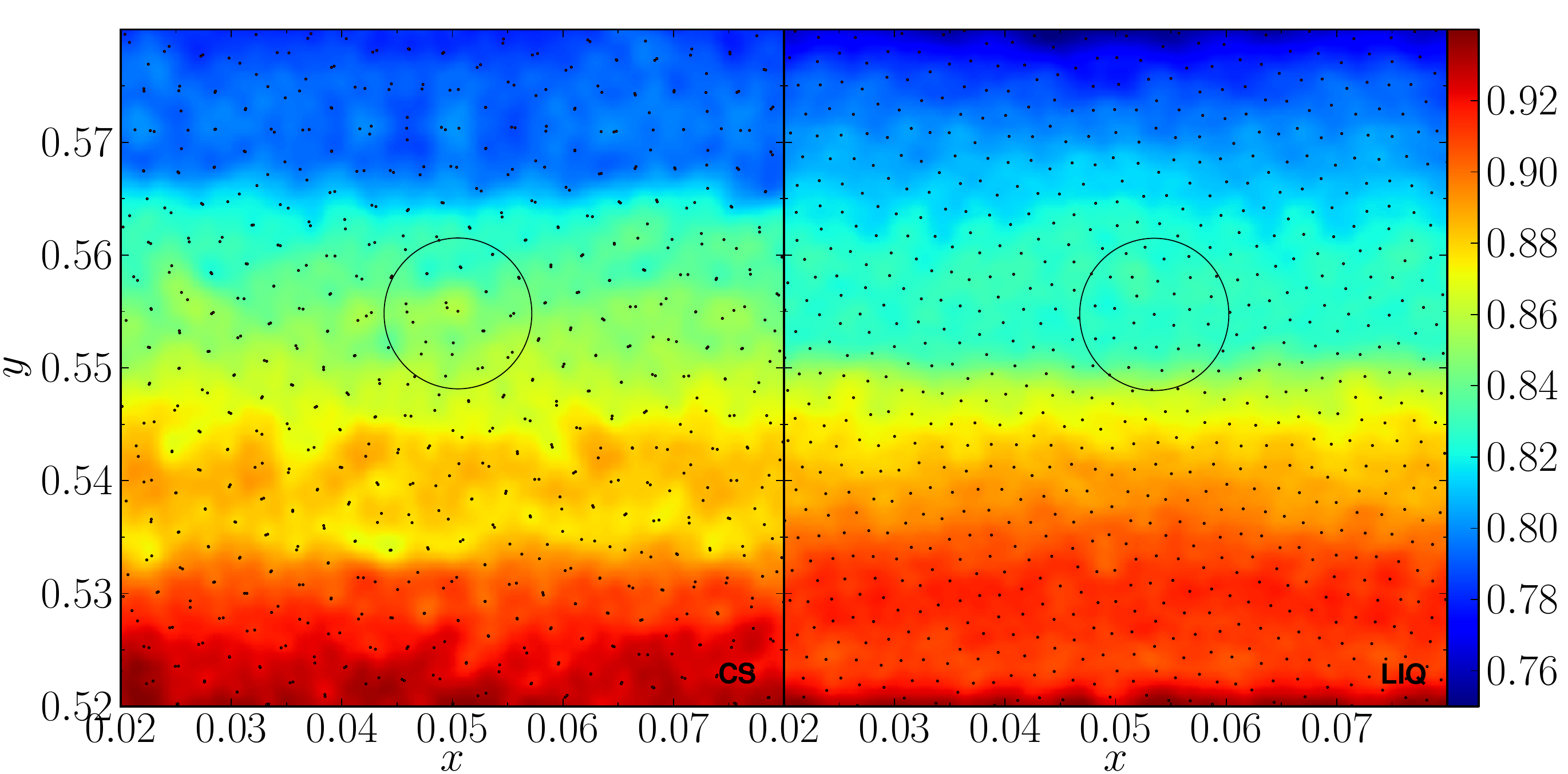}
  \caption{Particle distribution on top of the rendered density for a
    region in the shock tube test. {\em Left panel}: Cubic
    Spline. {\em Right panel}: LIQ Kernel. The overplotted circle is
    the region of influence for the same particle in both runs. The
    particle clumping in the left panel, using the CS kernel, is
    striking. No clumping is seen in the right
    panel.\label{fig:shockTubeRender}}
\end{figure}

In Fig. \ref{fig:shockTubeRender} we show the particle distribution on
top of the rendered density, for a small region of the shock tube, at
$t=0.2$. Overplotted is the circle of interaction for the same SPH
particle. The clumping behavior in the left plot, using the CS kernel,
is striking: particles form groups of 2. This behavior has been known
for some time \citep[see e.g.][]{schuessler1981}.  When using the LIQ
kernel (right panel) there is no clumping. We note that this clumping
does not lead to an actual decrease in resolution, as the circle of
interaction for the SPH particle has a comparable radius in the two
plots. Indeed, the clumping does not affect the actual size of the
smoothing region. It does affect the homogeneity of the particles in
that region, and through that it affects the validity of using an SPH
estimate of a continuous integral (see section \ref{sec:theCode}). As
the CS kernel has been used extensively in SPH codes and found to give
excellent results \citep[see e.g.][the shock tube results in
  Fig. \ref{fig:SodShockTube}]{springel2005, price2008, kawata2009},
the SPH estimate of the integral does not seem to suffer from the
clumping.


\begin{figure}
\includegraphics[width=\hsize]{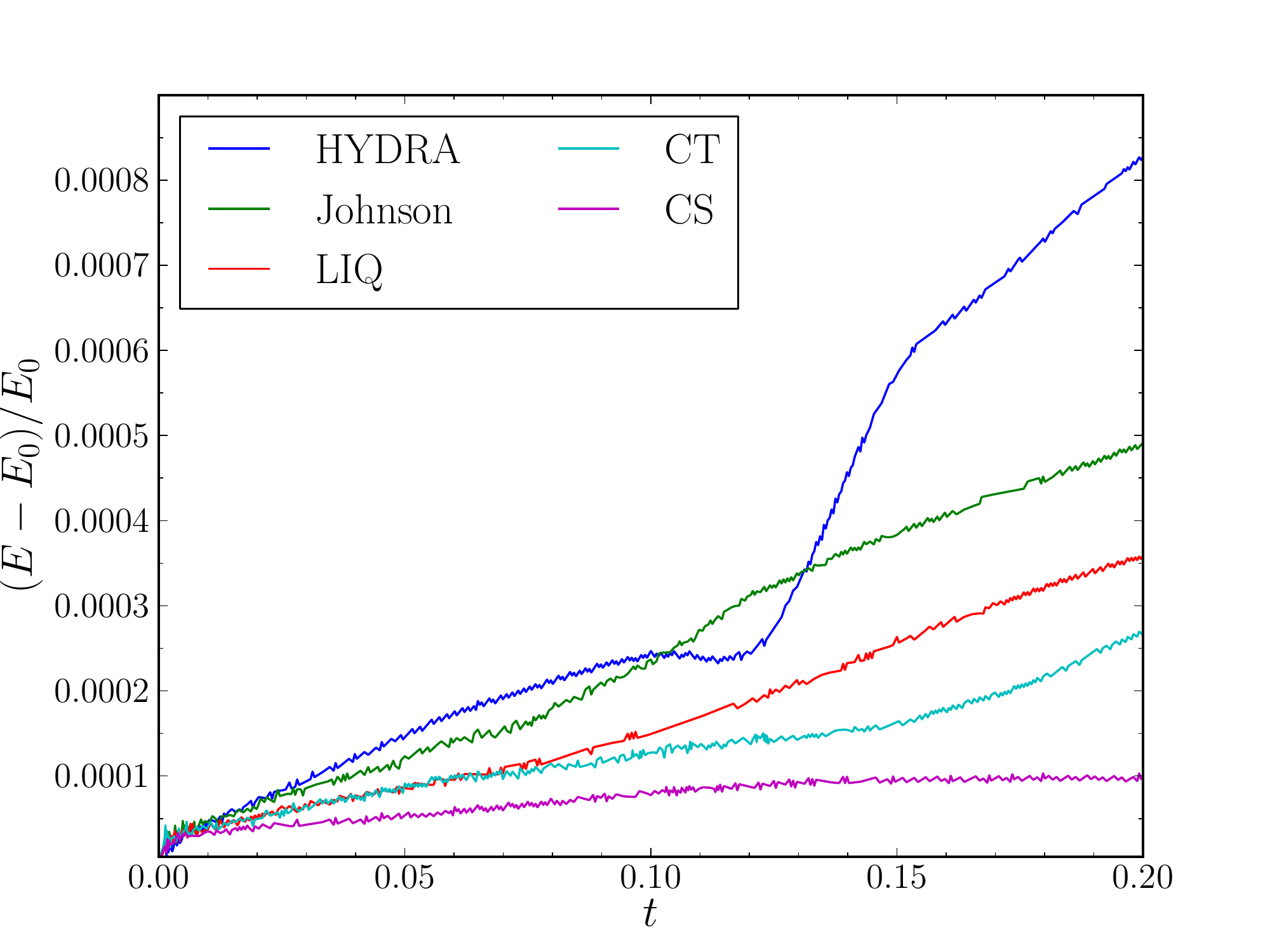}
\caption{Relative energy error as a function of time, for a number of
  2D shock-tube tests, using different smoothing kernels. Lines
  indicate, from top to bottom: HYDRA kernel (CS with modified central
  derivative), Johnson kernel (quadratic), LIQ kernel (see section
  \ref{sec:liqKernel}, linear + 4$^{\text{th}}$ order), CT kernel
  (centrally modified CS), Cubic Spline kernel (3$^{\text{rd}}$
  order). Note how modifying the central value of the derivative of
  the CS kernel drastically reduces energy
  conservation.\label{fig:st_energy}}
\end{figure}

In Fig. \ref{fig:st_energy} we show the energy conservation of SPH for
the 2D Sod shock-tube test for the HYDRA, Core Triangle, Johnson, LIQ
($x\rs{s}=0.3$) and Cubic Spline kernels. As expected, the Cubic Spline
kernel clearly leads to the best energy conservation. The behavior of
the CT and LIQ kernels is similar, especially in their initial
evolution. The Johnson and HYDRA kernels show the largest growth in
their respective relative energy errors. Note that the HYDRA kernel is
identical to the CS kernel, apart from the modified central first
derivative. The resulting kernel does prevent clumping but the energy
conservation leaves much to be desired. Moreover, among the family of
LIQ kernels, the choice $x\rs{s}=0.3$ turned out to provide the best
energy conservation. The different result for the LIQ and CT kernels
arises between $t=0.07$ and $t=0.15$. This interval corresponds to the
time where the initial rectangular symmetry in the simulation breaks
up. The initial energy error ($t < 0.07$) thus shows the performance of
the SPH scheme on a regular particle distribution whilst the late energy
error ($t > 0.15$) shows the performance with SPH particles having
arranged themselves according to the respective smoothing kernels. The
poor performance of the HYDRA kernel can be attributed to the artificial
modification of the central derivative of the kernel, which breaks the
conservative form of SPH. Note that all simulations use individual
time-steps which in itself breaks the time-symmetry (and hence the
energy conservation) of the integrator.

%% file: shearingLayers.tex
\section{Shearing Layers Test}
\label{sec:shearingLayers}
The shearing layers test, as presented by \citet{agertz2007}, is an
excellent way of examining the ability of a hydrodynamical code to
capture the formation of Kelvin-Helmholtz instabilities. These
instabilities occur when two fluids have a relative velocity
perpendicular to their contact layer. For an incompressible fluid
there is an analytical expression for the time-scale for the growth of
these Kelvin-Helmholtz instabilities \citep{chandrasekhar1961, price2008}:
\begin{equation}
  \tau\rs{KH} = \frac{2 \pi}{\omega},
  \label{kh_timescale}
\end{equation}
where
\begin{equation}
  \omega = \frac{2 \pi}{\lambda} \frac{(\rho_1\rho_2)^{\frac{1}{2}}|v_{x,1} - v_{x,2}|}{(\rho_1 + \rho_2)}.
\end{equation} 
The indices 1 and 2 denote the two fluid layers. $\lambda$ is the
wavelength of the perturbation. When re-written in terms of the
density contrast $\chi$ ($\rho_1 = \chi \rho_2$, we make the
convention $\chi \geq 1$ e.g. the index 1 denotes the high-density
layer):
\begin{equation}
  \omega = \frac{2 \pi}{\lambda} \frac{(\chi)^{\frac{1}{2}}|v_{x,1} - v_{x,2}|}{(1 + \chi)}.
  \label{eq:omega_chi}
\end{equation}
Various approaches have been suggested to enforce incompressibility in
SPH \citep[e.g.][]{cummins1999, hu2009}. We use the weakly
compressible SPH formulation, approximating incompressibility by using
a small Mach number. Note that incompressibility is a requirement only
for equation (\ref{eq:omega_chi}) to be valid, it is not a requirement
for the simulations themselves to be valid.

KH instabilities are seeded by introducing an initial $y$-velocity
perturbation following the prescription:
\begin{equation}
  v_y = A \sin(2\pi x / \lambda).
\end{equation}
All simulations, unless noted otherwise, have the following setup: the
$x$- and $y$ ranges are 1. For the adiabatic index $\gamma$ we use a
value of 5/3, as applicable for an ideal monoatomic gas.  The density
contrast $\chi$ is 10, with low- and high-density layers having a
density of respectively 1 and 10. The respective specific energies are
15 and 1.5, resulting in respective sound speeds of 4.08 and
1.29. $\lambda$ is fixed at $1/6$, which should result in the initial
growth of 6 KH density rolls. $A$ is set to 0.025. Note that we focus
on a density contrast of 10, not 2, because preliminary tests showed
us that an SPH code has much less trouble reproducing Kelvin-Helmholtz
instabilities for lower density contrast: we concentrate on the more
demanding test here. Previous studies of the KH instabilities either
restrict themselves to a density contrast of 2 \citep{read2010,
  merlin2009} or show KH tests with KH time-scales of 0.6 or lower
\citep{price2008, kawata2009}.

\input{gridcodesection}
\begin{figure*}%
\includegraphics[width=\hsize]{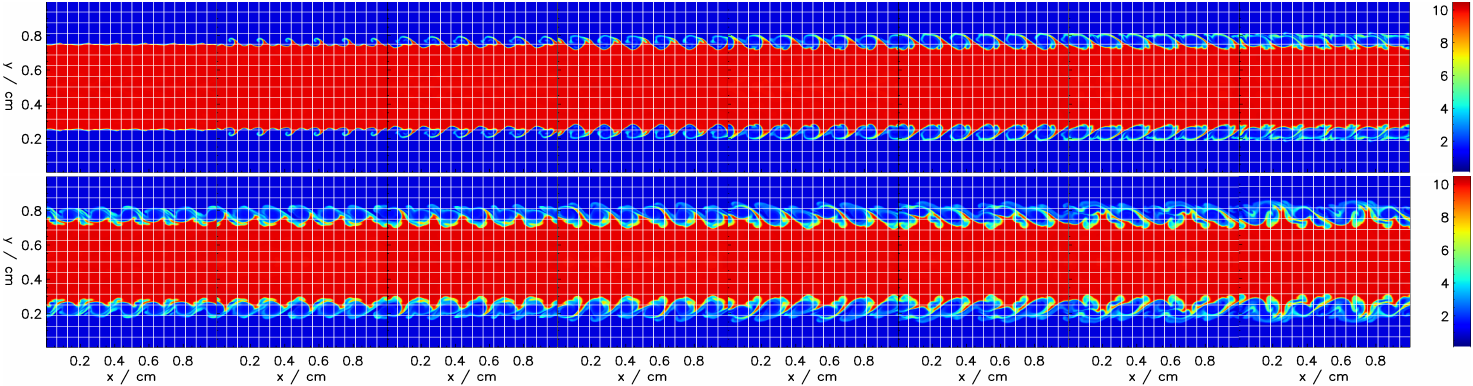}%
\caption{Grid simulation of the shearing layers test with $M=0.2$. Top
  row, left to right: $t=0.25, 0.5, 0.75, 1, 1.25, 1.5, 1.75,
  2$. Bottom row, left to right $t=2.25, 2.5, 2.75, 3, 3.25, 3.5,
  3.75, 4$: White boxes mark blocks of $16^2$ grid
  cells.\label{fig:elke_kh_02}}
\end{figure*}
\begin{figure*}%
\includegraphics[width=\hsize]{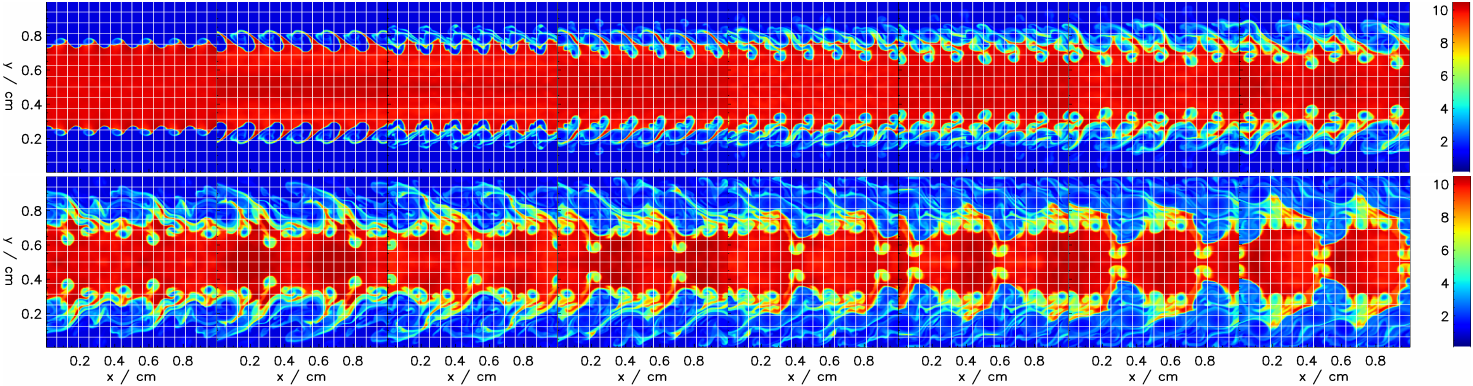}%
\caption{Grid simulation of the shearing layers test with $M=0.6$. Top
  row, left to right: $t=0.25, 0.5, 0.75, 1, 1.25, 1.5, 1.75,
  2$. Bottom row, left to right $t=2.25, 2.5, 2.75, 3, 3.25, 3.5,
  3.75, 4$: White boxes mark blocks of $16^2$ grid
  cells.\label{fig:elke_kh_06}}
\end{figure*}

\subsection{Standard SPH}
\label{subsec:standardSPH}
\begin{figure*}
\includegraphics[width=\hsize]{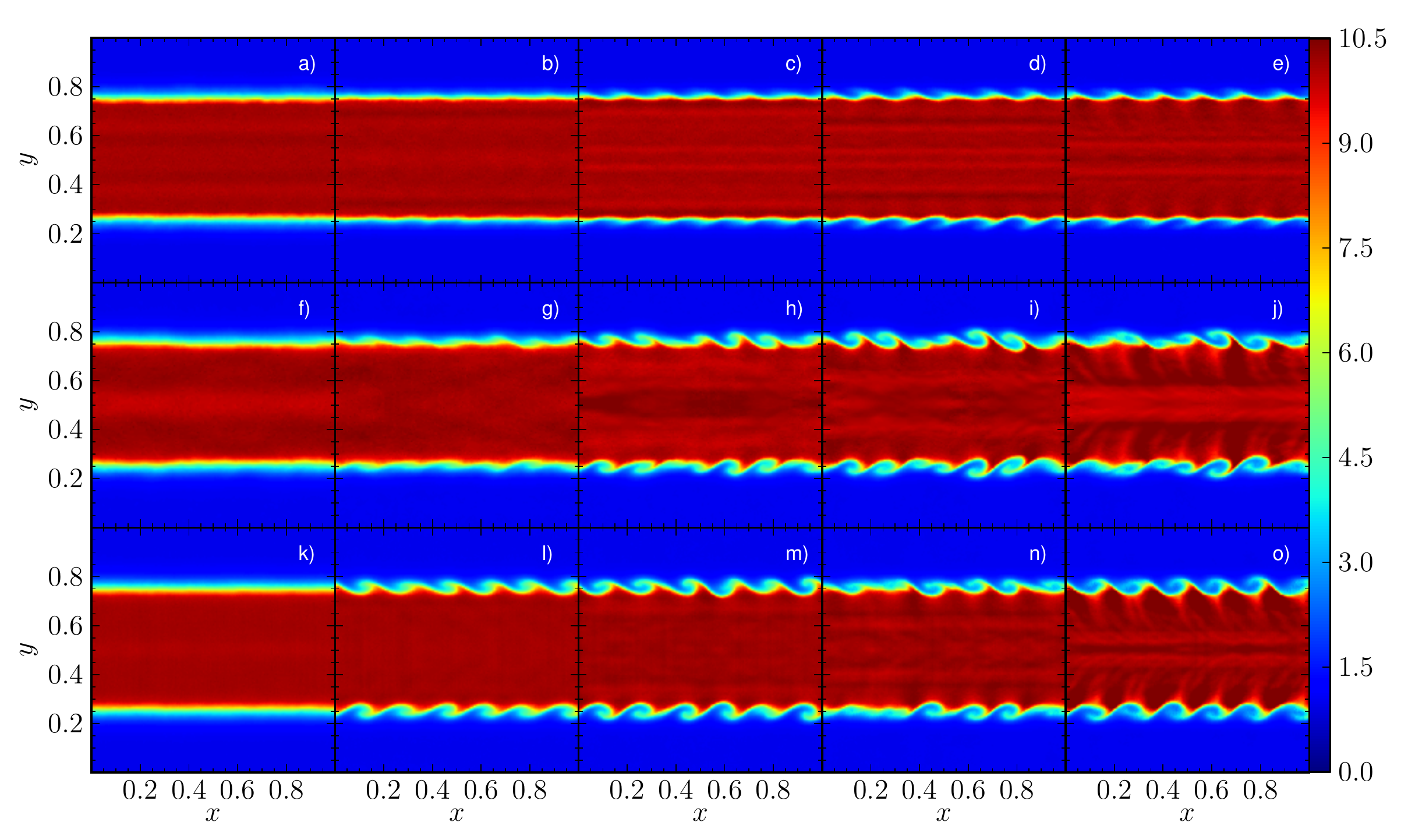}
\caption{Density plots for shearing layers simulations (table
  \ref{tab:khsims2}) at their respective $\tau\rs{KH}$ (table
  \ref{tab:khsims}). {\em Top row (a-e)}: from left to right:
  simulations SPH1-SPH5: standard SPH simulations. {\em Middle row
    (f-j)}: RHO1-RHO5: standard SPH with a smoothed density
  setup. {\em Bottom row (k-o)}: LIQ1-5: SPH with a smoothed density
  and using the LIQ kernel.\label{fig:sph_KH}}
\end{figure*}

\begin{figure*}
\includegraphics[width=\hsize]{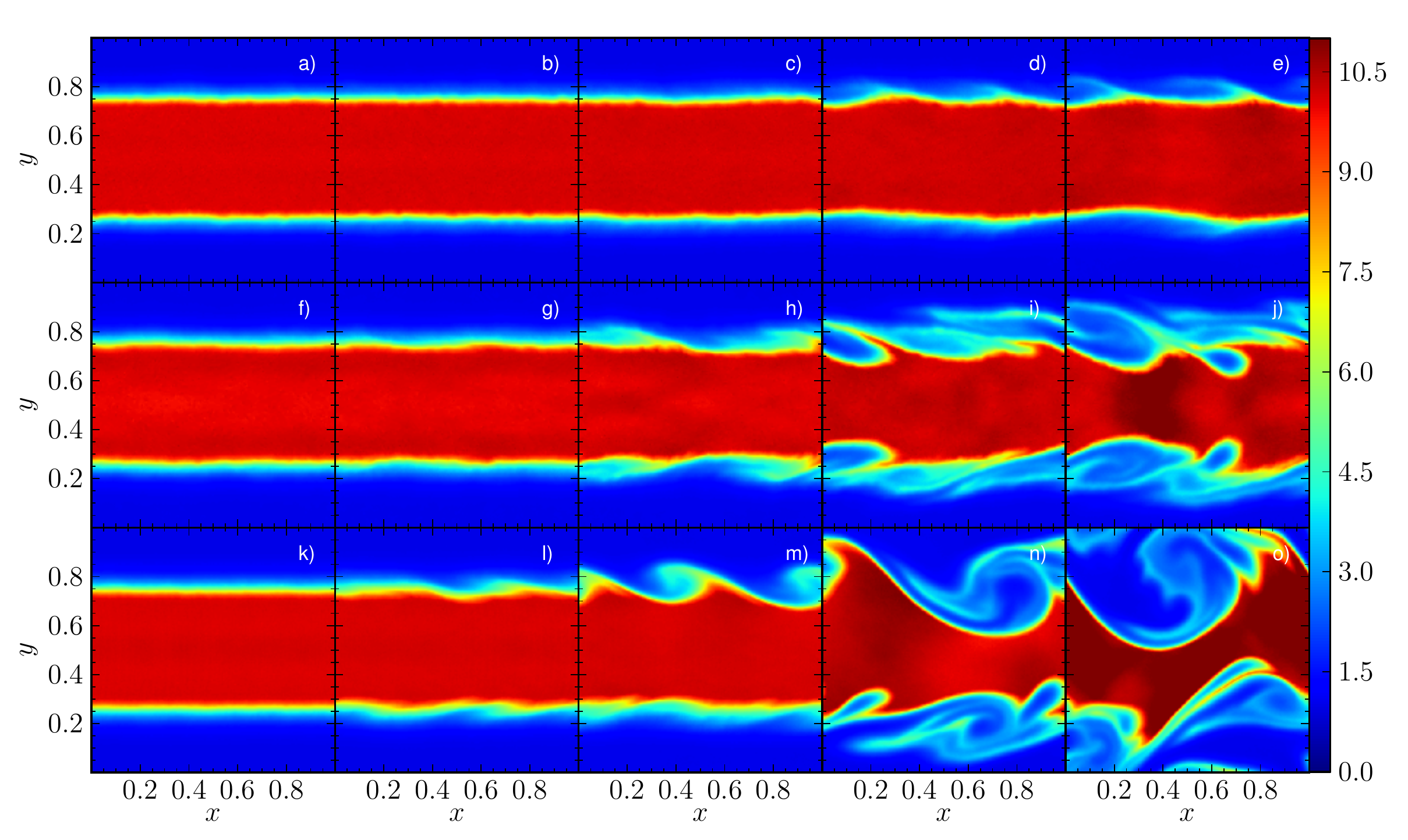}
\caption{Same figure as Fig. \ref{fig:sph_KH}, all figures now at
  $t=2$.\label{fig:sph_KH_2}}
\end{figure*}

We first examine the capability of standard SPH to handle the shearing
layers problem. With standard SPH we mean SPH as implemented by
e.g. \citet{springel2005} (which uses the entropy formulation of SPH)
{\em augmented with the artificial conductivity formalism} (see
section \ref{sec:theCode}). This form of the SPH equations is rapidly
becoming the standard \citep[see e.g.][]{price2008, kawata2009,
  vanaverbeke2009, merlin2009}. We do this test for a range of Mach
numbers, and thus for a range of Kelvin-Helmholtz time-scales
(eq. (\ref{kh_timescale})). The specific energies of the SPH particles
were set after the density is computed, in order to force perfect
pressure equilibrium (thus applying a crude smoothing of the
IC). Details for the different simulations are listed in tables
\ref{tab:khsims} and \ref{tab:khsims2}. Rendered plots of the density
are shown in Fig. \ref{fig:sph_KH}, for each run at its respective
$\tau\rs{KH}$ (see table \ref{tab:khsims2}). From
Fig. \ref{fig:sph_KH} we learn that changing the Mach number in the
simulation has a drastic impact on the formation of KH rolls: for
small $M$ they are completely absent. The simulations of
\citet{price2008} with $\chi=10$ have $M=0.775$ for the high-density
layer, they are to be compared with SPH4. We thus find that the
results in \citet{price2008}, where artificial conductivity was enough
to enable SPH to produce KH rolls, are only valid for high enough Mach
numbers. In following sections we will try to improve on this
situation.  The situation at $t=2$ is shown in
Fig. \ref{fig:sph_KH_2}. It tells us that there is a large difference
in the long-term evolution of the shearing layer simulations depending
on the setup and the ingredients of the code. On the panels for
$M=0.6$ (c, h and m) we should see $\lambda=1/2$ instabilities when
comparing with the grid simulations in
Fig. \ref{fig:elke_kh_06}. Panel h shows a hint of these instabilities
but they are very poorly developed. Panel m does better, the KH rolls
appear although only on one side of the high-density layer.
\begin{table}
  \begin{minipage}{60mm}
  \caption{Velocities and $\tau\rs{KH}$ for the different KH
    simulations. Rows from top to bottom: suffix: simulation number
    suffix (e.g. for the SPH series: SPH1-SPH5), $M$: Mach number of
    the high-density layer, $|v_x|$: absolute value of the
    $x$-velocity, $\tau\rs{KH}$: KH time-scale
    (eq. (\ref{kh_timescale}))\label{tab:khsims}}
  \begin{tabular}{cccccc}
    suffix        & 1 & 2 & 3 & 4 & 5 \\
    \hline\hline
    $M$           & 0.2  & 0.4  & 0.6  & 0.8  & 1    \\
    $|v_x|$       & 0.26 & 0.52 & 0.77 & 1.0  & 1.3  \\
    $\tau\rs{KH}$ & 1.23 & 0.56 & 0.37 & 0.28 & 0.22 \\
    \hline
  \end{tabular}
  \end{minipage}
\end{table}
\begin{table}
  \begin{minipage}{82mm}
  \caption{Parameters for the different KH simulations. {\em series}:
    Name for the simulation series, e.g. SPH1--SPH5. {\em \# part}:
    Number of particles used. {\em AC}: Artificial conductivity
    included? (yes: X). {\em kernel}: Smoothing kernel used. {\em
      $\rho$--smoothing}: Type of density smoothing applied, if
    any.\label{tab:khsims2}}
  \begin{tabular}{cccccc}
    series            & SPH    & RHO    & LIQ     & GRID   & NOAC \\
    \hline\hline
    \# part           & 199252 & 184100 & 184100 & 200788 & 184100  \\
    AC                &   X    &   X    &   X    &   X    &   -     \\
    kernel            &   CS   &   CS   &   LIQ  &  CS/LIQ&   LIQ    \\
    $\rho$--smoothing &   -    & column & column & grid   & column  \\
    \hline
  \end{tabular}
  \end{minipage}
\end{table}
\begin{figure*}
\includegraphics[width=\hsize]{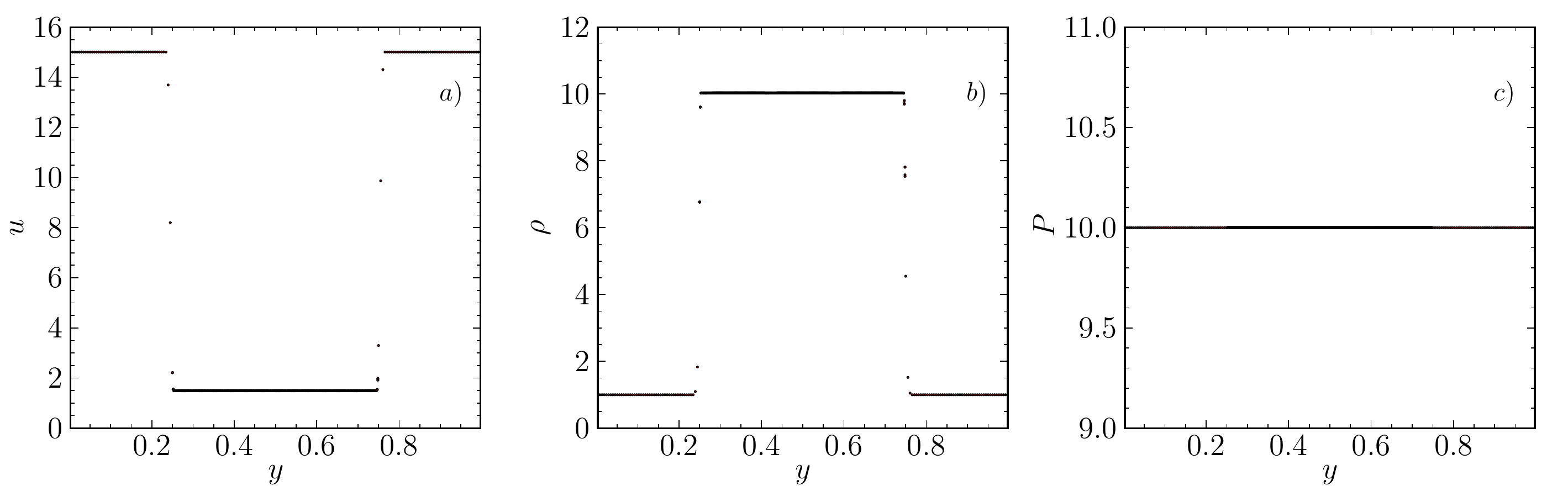}
\caption{Initial specific energy (a), density (b) and pressure (c) as
  a function of $y$ for the SPH1-5 (see table \ref{tab:khsims})
  simulations. Only particles in the $x$-interval [0,0.005] are
  shown. Note that the specific energy of the particles at the contact
  discontinuity is smoothed in order to avoid a pressure
  blip.\label{fig:standardSPH_IC}}
\end{figure*}
\subsection{Standard SPH - Column-smoothed IC}
\label{subsec:smoothIC}
We can expect the way the initial conditions for the shearing layers
test were set up in the previous section to cause problems (as briefly
mentioned by \citet{hess2009}, though they do not try to formulate a
remedy). SPH is inherently smoothed, and, in the current formulation,
this particle smoothing is isotropic as the smoothing length is a
scalar, not a vector nor a tensor. When one uses this formulation of SPH
to tackle a problem where sharp discontinuities are present, as is the
case in the shearing layers test, one is forcing SPH into a situation
where its behavior is not well defined. Smoothing the initial particle
energy based on the calculated initial density and a requested constant
pressure value does not remedy this problem: it enforces equal pressures
on a particle basis, which does not necessarily result in pressure
equilibrium on a simulation basis. This is easy to see when making an
analogy with the density: giving all particles an equal mass results in
a constant density only if the particle distribution is
homogeneous. \citet{robertson2010} show that Eulerian codes exhibit
similar problems, with numerical diffusion wiping out small-scale
structures in the presence of a large bulk flow and sharp
discontinuities. They present a different version of the KH test which
does not suffer from this. We however do not discard the canonical test
because the code is not able to solve it. We choose to modify the test
so that we retain the physics of the original test whilst allowing the
code to deal with it

\begin{figure*}
\includegraphics[width=\hsize]{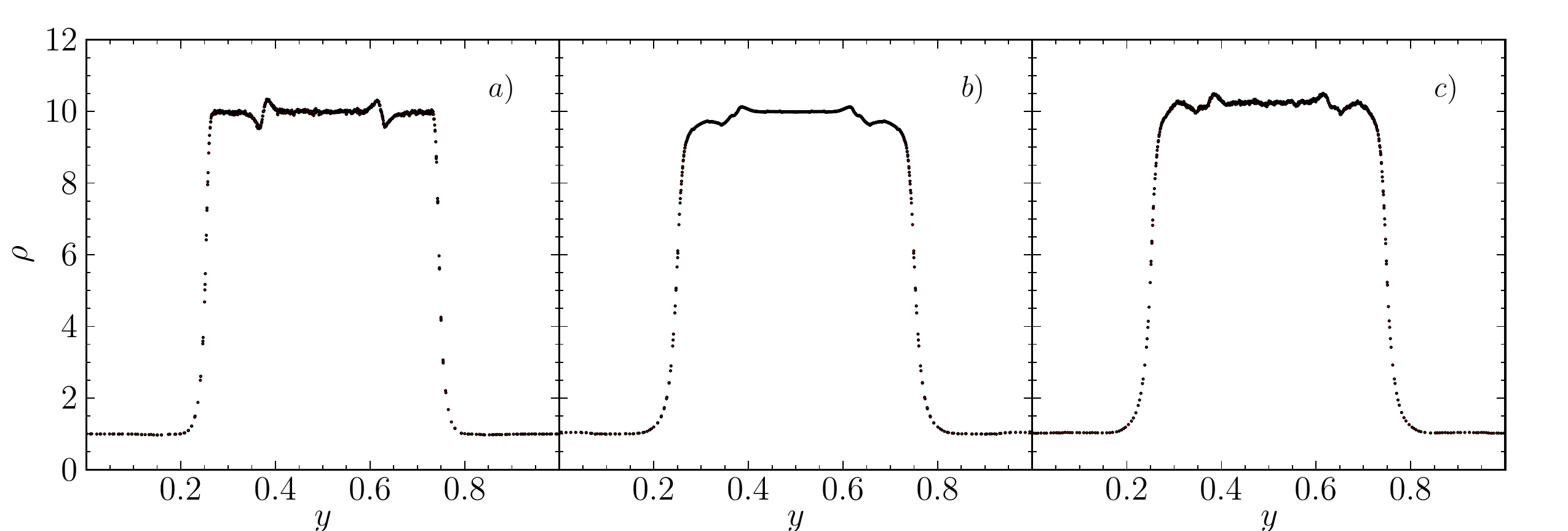}
\caption{{\em panel a}. simulation SPH1 at $t=0.1$ (particles limited
  to $x$-interval [0,0.005]). One clearly sees shocks travelling
  inward from the contact discontinuities. {\em panel b}. simulation
  RHO1. Shocks are still present, although reduced in magnitude. {\em
    panel c}. density smoothing + the LIQ
  kernel.\label{fig:sph_shocks}}
\end{figure*}

Artificial conductivity goes a long way in {\em preventing}
discontinuities as sharp as these from arising in simulations, but by
the time AC has sufficiently smoothed the initial discontinuity a series
of {\em shocks} has been created in the simulation volume. These are
shown in Fig. \ref{fig:sph_shocks}. Depending on the time-scale of the
growth of the instabilities these shocks will play a role in destroying
emerging KH instabilities: the longer it takes for the instabilities to
manifest themselves, the more they are wiped out by the travelling shock
waves. When the shocks pass through the emerging instabilities a mixed
layer is formed at the contact discontinuity. This layer acts as a
lubricant, separating the two layers and thus preventing KH
instabilities from developing. Only in the case where the KH
instabilities form fast enough compared to the scale on which they are
destroyed are they able to survive. This can be translated into a
general paradigm: {\em the longer it takes for KH rolls to manifest
themselves in SPH, the more time there has been for SPH-induced
deviations from the analytical, theoretical problem to destroy
them}. This explains why it is more easy to produce KH rolls for high
Mach numbers. The origin of these shocks is probably related to the LMI,
in the same way as the artificial gap problem. Imagine two particles
with equal pressures, but different densities. If these particles
approach each-other their respective densities will change towards
equality. As entropy is conserved this implies, with $p=A \rho^\gamma$
\citep{springel2002}, that these particles now have different
pressures. As this process happens everywhere along the initial contact
discontinuity this can set a shock-wave in motion. A possible cure for
this problem is to smooth the regions where these discontinuities occur,
giving rise to a net smaller LMI effect. To create a smooth interface we
choose a function of the form:
\begin{equation}
  \rho(y) = A\ \mathrm{atan}(B(y+C)) + D,
  \label{eq:boundFitFunc}
\end{equation}
where
\begin{eqnarray}
  A &=& \frac{\rho_1 - \rho_0}{2\ \mathrm{atan}(\beta)} \\
  B &=& 2 \frac{\beta}{\delta} \\
  C &=& -\frac{\delta}{2} \\
  D &=& \frac{\rho_1 + \rho_2}{2}
\end{eqnarray}
were fixed requiring (\ref{eq:boundFitFunc}) to be symmetrical around
the boundary, scaling the argument of the $\mathrm{atan}$ such that
$y=0, \delta$ resulted in an argument of respectively $-\beta,
\beta$. For $\beta$ a value of 10 gives reasonable results, because
$\atan(10)$ is close to $\pi/2$ (which is needed because it is
connected to two constant-density functions). $\delta$ is the width of
the boundary, which we took to be $(y\rs{max} - y\rs{min}) / 7.5$ for
a part of the simulation box containing a single boundary layer
e.g. $\delta=0.5/7.5$ in the current setup.

There are several methods to set up a smooth density. One method is to
set up the particles in the two regions on two different grids. Within
such a grid the $x$- and $y$ separations ($\Delta x$ and $\Delta y$
respectively) of particles are equal. Denoting one layer with the
subscript 1, the other with 2, we have of course $\Delta x_1 \ne
\Delta x_2$ and $\Delta y_1 \ne \Delta y_2$ when the layer densities
differ. The smooth boundary between those grids can then be
constructed by putting a number of equal-$y$ rows between those
grids. The distance between these rows varies from $\Delta y_1$ to
$\Delta y_2$ according to the square (in the 2D case) of some chosen
function (e.g. eq. (\ref{eq:boundFitFunc})). The number of particles
on a row is then fixed by comparing the computed SPH density of
particles on that row to the value of required analytical density
$\rho\rs{an}$ at that point. To get a perfect representation of the
bounding function one would have to use an iterative scheme, varying
the number of particles on the rows, recomputing the smoothing lengths
and densities and repeating until convergence is reached. In our setup
our prime concern is that the density is smooth, it is of little
importance if the analytical boundary function is perfectly
reproduced. We thus use a shortcut in generating the initial
conditions: the SPH density on the $i$-th row $\rho_i$ is esimated by:
\begin{equation}
  \rho_i = \frac{m_i}{\Delta x_i \Delta y_i},
  \label{eq:sphDensityEstimate}
\end{equation}
with $m_i$ the particle mass on that row, $\Delta y_i$ the distance to
the previous row and $\Delta x_i$ the interparticle distance. $\Delta
x_i$, and from that the number of particles, can then be found from:
\begin{equation}
  \Delta x_i = \frac{m_i}{\rho\rs{an} \Delta y_i}.
\end{equation}
Another method to construct a smooth density is to put all the SPH
particles on columns (lines with constant $x$), with an equal and
unchanged intercolumn distance. The smoothed density can then be
attained by varying the interparticle separation within those
columns. Finding those separations, using equation
\ref{eq:sphDensityEstimate}, is then a simple matter of using some
bisection algorithm. 

Simulations RHO1-5 correspond to the SPH1-5 series, with their initial
density discontinuity smoothed using the first smoothed density
method: particles on columns. Results are shown in
Fig. \ref{fig:sph_KH}. When comparing the top and middle rows we see
that for $M \ge 0.6$ KH rolls have appeared. For $M=0.4$ some bumps
are present, but these do not evolve into KH rolls. Although we have
already greatly improved the situation, for low Mach numbers we are
still unable to get adequate results. In Fig. \ref{fig:sph_shocks} we
can see that the magnitude of the shocks has decreased because of the
density smoothing.
%
%
%
\subsection{Modified kernel - Column-smoothed IC}
As mentioned in section \ref{sec:liqKernel} we can expect to do
better in very sensitive simulations when using a smoothing kernel
that does not cause particle clumping. In Fig. \ref{fig:sph_KH} we
show results for the same simulations as the RHO1-RHO5 simulations
(e.g. with smoothed density), but now using the LIQ kernel (see
section \ref{sec:liqKernel}). We see KH instabilities appear for $M
\geq 0.4$, which is an improvement over the previous results where
clear KH rolls appeared for $M \geq 0.6$. We also see that in general
the shape of the KH rolls is much more symmetrical.

\begin{figure}
\includegraphics[width=\hsize]{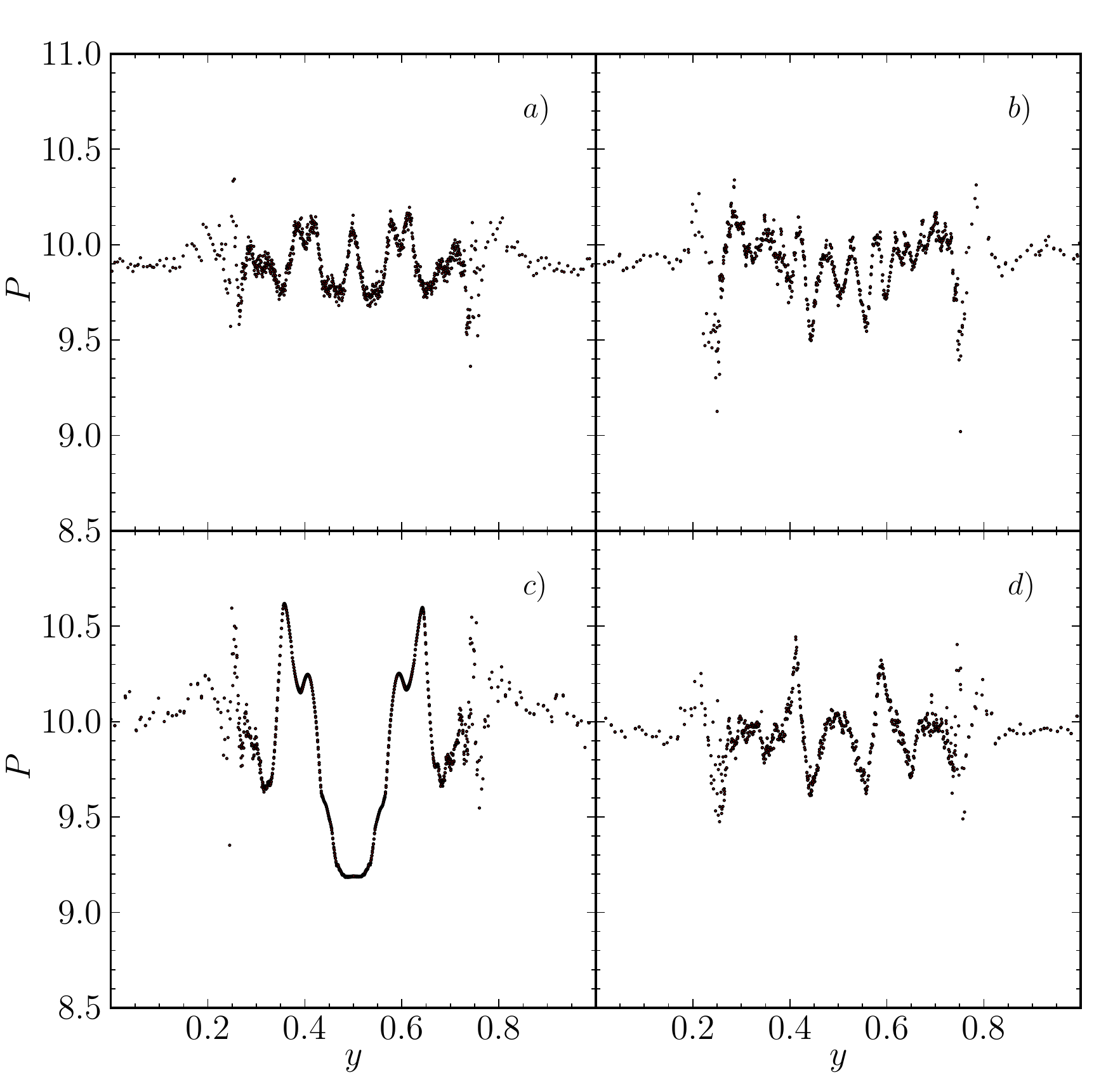}
\caption{Pressure of SPH particles in the $x$-interval [0,0.005] as a
  function of $y$, at $t=0.25$, with $M=0.4$. The top row shows two
  runs with LIQ2 setup. The bottom row shows RHO2. The left panels
  have column density smoothing, the right panels have grid-based
  density smoothing. \label{fig:pressureShocks}}
\end{figure}

In figure \ref{fig:pressureShocks} we show the pressure of the
particles at roughly half the KH time-scale. When we look at the column
smoothing results it is clear that the dominant shock in the RHO2
simulation, around $y=0.5$, is greatly reduced in magnitude in the LIQ2
simulation. This difference is mainly due to the different way the
kernels respond to the initial conditions (particles initially on
columns): the CS kernel is clearly not a good choice in this case. This
is also demonstrated in the bottom right panel of
Fig. \ref{fig:pressureShocks}, where a result for the CS kernel using
grid smoothing is shown. It compares favourably with both LIQ2 results.

%
%
\subsubsection{Particle number}
\begin{figure}
\includegraphics[width=\hsize]{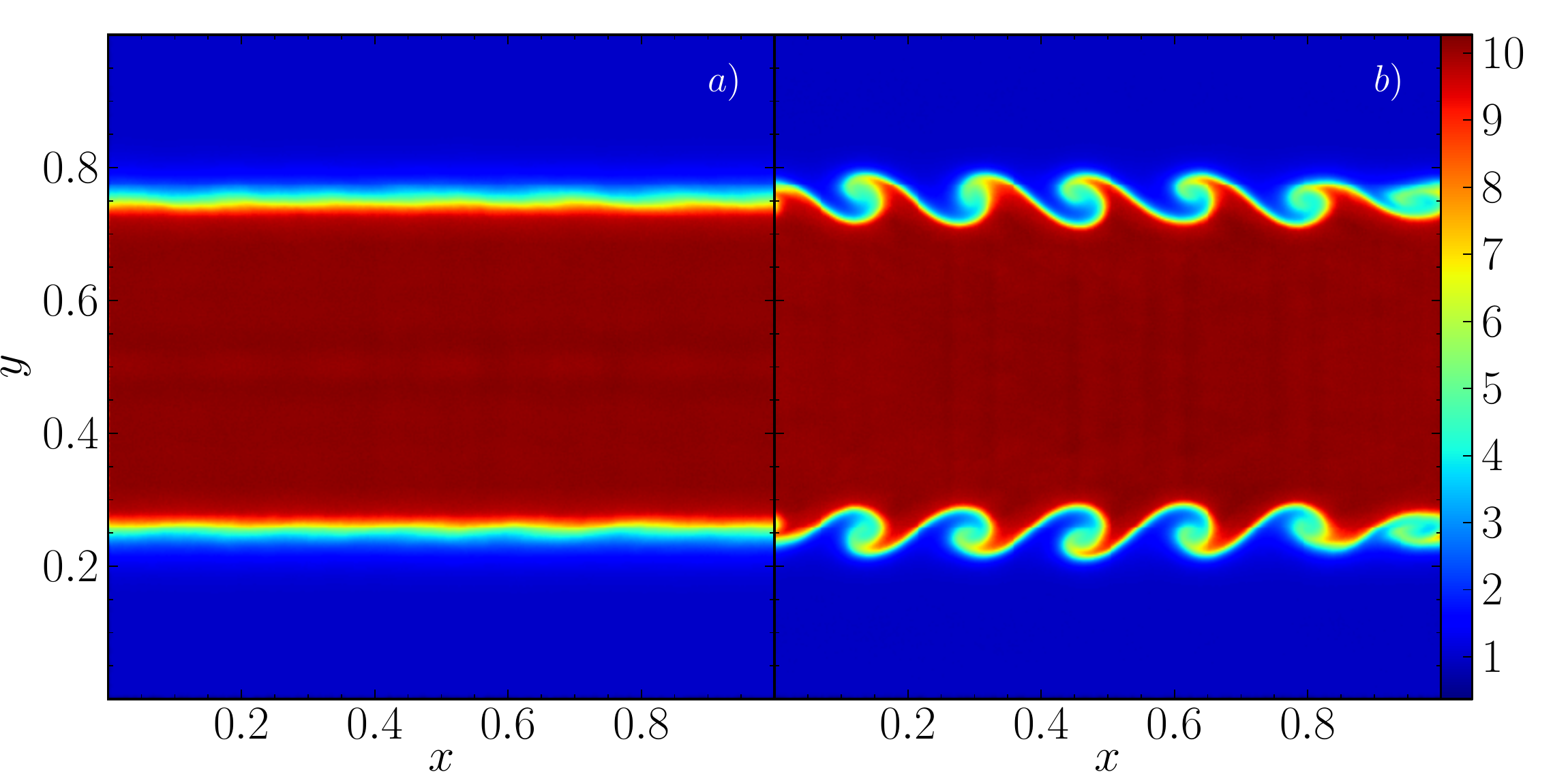}
\caption{Density of two increased resolution shearing layers
  simulations at their respective $\tau\rs{KH}$. {\em panel a}. LIQ+1
  ($M=0.2, 551\,100$ particles). {\em panel b}. LIQ+2 ($M=0.4,
  551\,100$ particles). \label{fig:LIQ_particle}}
\end{figure}

We can try increasing the particle number for the simulations with low
Mach number, to see how it affects the KH rolls: simulations LIQ+1 and
LIQ+2, each with $551\,100$ particles. Results are shown in
Fig. \ref{fig:LIQ_particle}. The LIQ+2 simulation, where KH rolls were
already present in its low-resolution counterpart LIQ2, is seen to
benefit a lot from the increased resolution: the KH rolls are much
better defined. The situation for LIQ+1 however has not changed: KH
rolls are still abscent at that resolution.
\subsection{Grid-based density smoothing}
\label{subsec:smoothIC2} 
\begin{figure}
\includegraphics[width=\hsize]{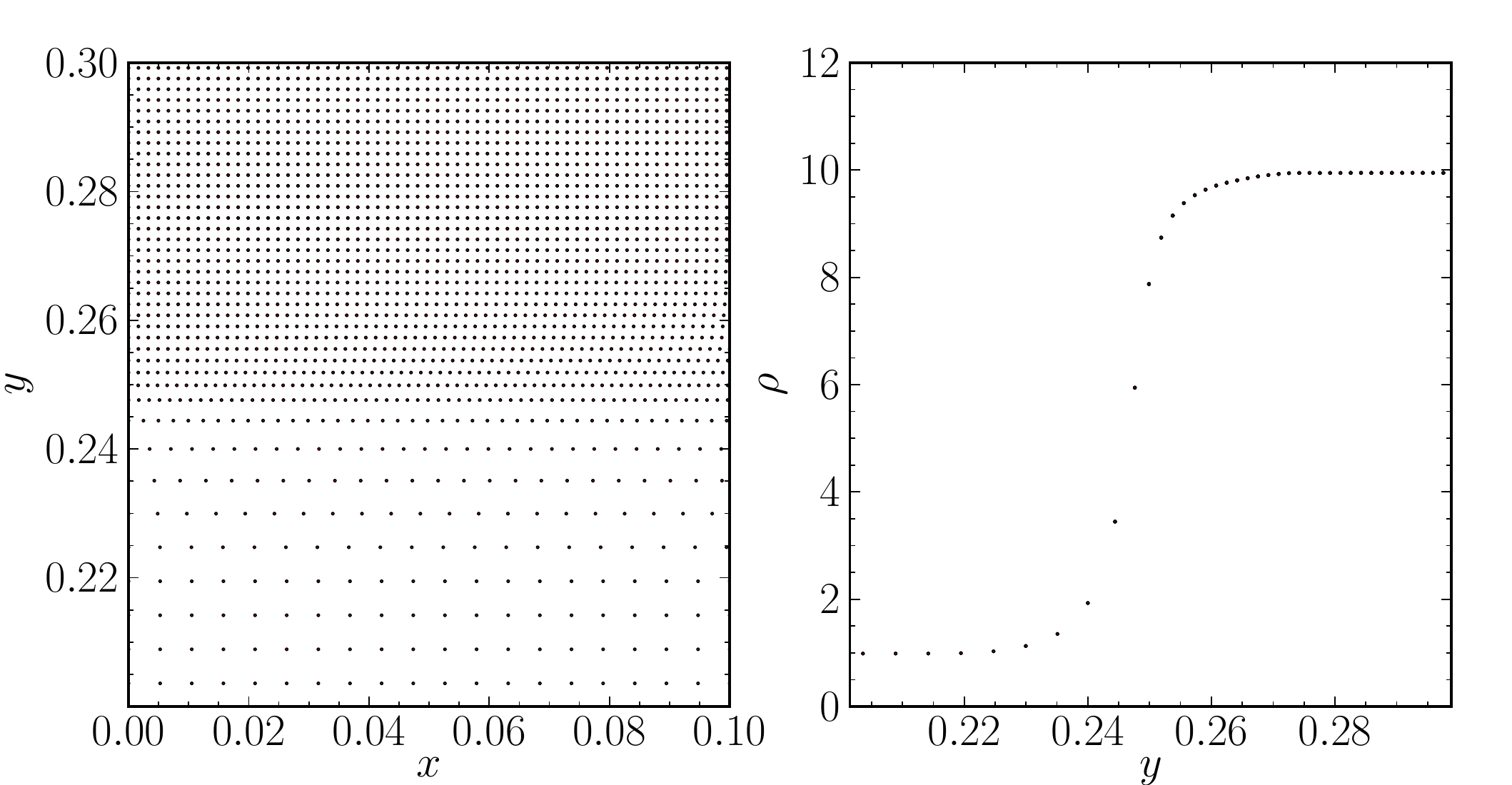}
\caption{{Initial conditions with smooth boundary from section
    \ref{subsec:smoothIC2} \em Left panel}: At the top and bottom we
  can see the high- and low density grids respectively. A smoothly
  varying layer lies in between. {\em Right panel}: SPH-computed
  density as a function of $y$, for the particles in the left
  panel. The density varies smoothly.\label{fig:smooth2}}
\end{figure}

\begin{figure}
\includegraphics[width=\hsize]{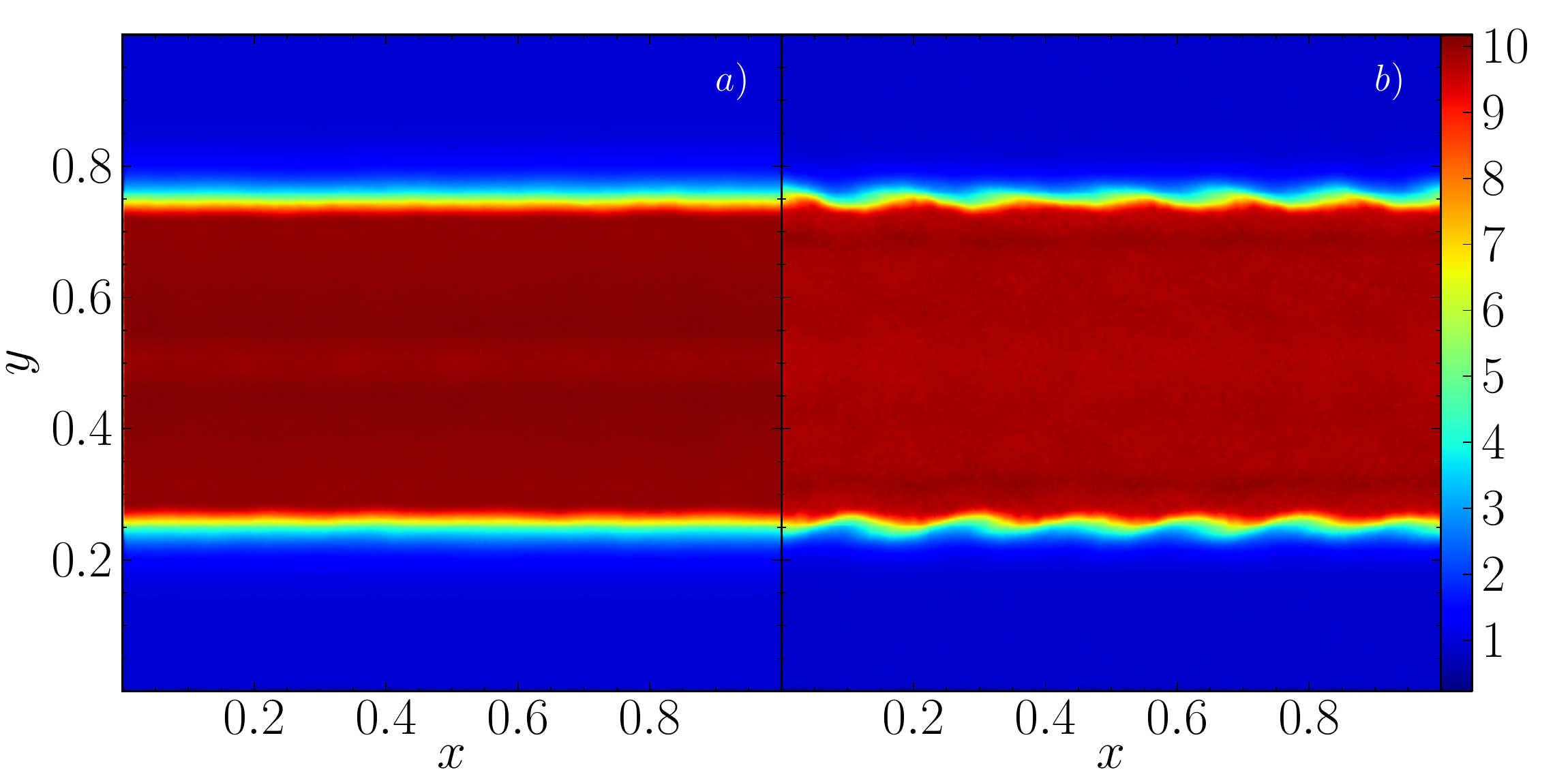}
\caption{Shearing layer simulations at their respective $\tau\rs{KH}$
  using grid-based density smoothing. {\em Left panel}:$M=0.2$, using
  the LIQ kernel. {\em Right panel}: $M=0.4$, using the CS kernel. No
  improvement is seen over the column-based smoothing results.
\label{fig:smoothGrid}}
\end{figure}

We have also constructed initial conditions
using the second smooth interface method: the two different layers
consist of two different grids, with two smooth interfaces at the
contacts. The energies of the particles are set using the same algorithm
as before: after the SPH densities are computed the particle specific
energies are set based on a constant pressure value. Part of the initial
particle distribution is shown in Fig. \ref{fig:smooth2}. The simulation
shown uses the LIQ kernel. Results for $M=0.2$ are shown in
Fig. \ref{fig:smoothGrid}. On the density plot we see that no KH rolls
manifest themselves: the same shocks that were present in the LIQ1
simulation are present here. We thus find that the artificial
conductivity is not able to smooth away initial energy discontinuities
fast enough to prevent the LMI from triggering these shock-waves. In the
right panel we show the rendered density of the RHO2 simulation, using
grid-based density smoothing (see also
Fig. \ref{fig:pressureShocks}). The observed ripples at the contact
discontinuities are a slight improvement over the RHO2 simulation shown
in figure \ref{fig:sph_KH}. The particle clumping does however prevent
the KH rolls from growing to the same size as their LIQ2 counterparts.

\begin{figure}
\includegraphics[width=\hsize]{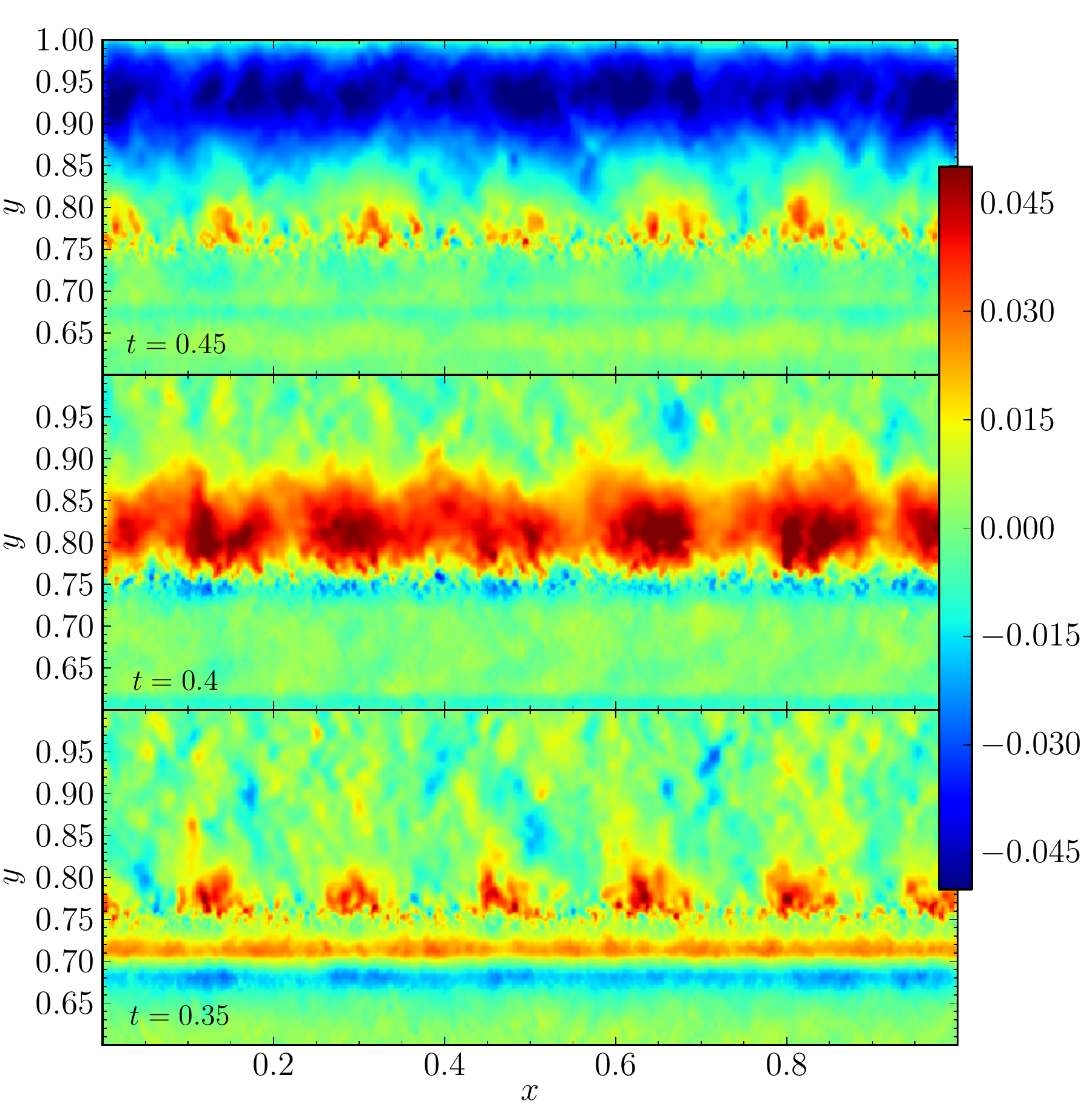}
\caption{Rendered $v_y$ for a simulation with $M=0.2$, LIQ kernel,
  grid-based density smoothing. From bottom to top: $t=0.35, 0.4,
  0.45$. {\em Bottom panel}: We see emerging KH instabilities as the 6
  red $v_y$ areas around $y=0.75$. Below them, around $y=0.7$, we have
  a shock wave travelling upwards, with a small layer of positive $v_y$
  in front of it and a region of negative $v_y$ in its wake. {\em
    Middle panel}: The shock is now almost at the same height of the
  instabilities. Its front is entering the low-density layer,
  increasing its $y$-length. {\em Upper panel}: The shock has passed
  through the instabilities. These now have a drastically reduced
  magnitude. The wake of the shock is still visible as the big blue
  band at the top of the panel.\label{fig:3406_vy}}
\end{figure}

In figure \ref{fig:3406_vy} we show the destruction of the KH
instabilities by a shock wave in detail. From bottom to top we can see
the wave passing through the emerging KH instabilities, erasing them.
\subsubsection{Relaxing}
Because we want to avoid any influence from shock waves, which we have
shown are moving through the simulation box, we set up a simulation
where we have first relaxed the initial conditions. The relaxing scheme
we used is a default SPH simulation, with the PdV terms removed from
the energy equation. The energy of particles is thus not allowed to
change due to expansion/contraction, only due to artificial viscosity
and conductivity. We use this scheme to stay as close as possible to
the original energy profile.

\begin{figure}
\includegraphics[width=\hsize]{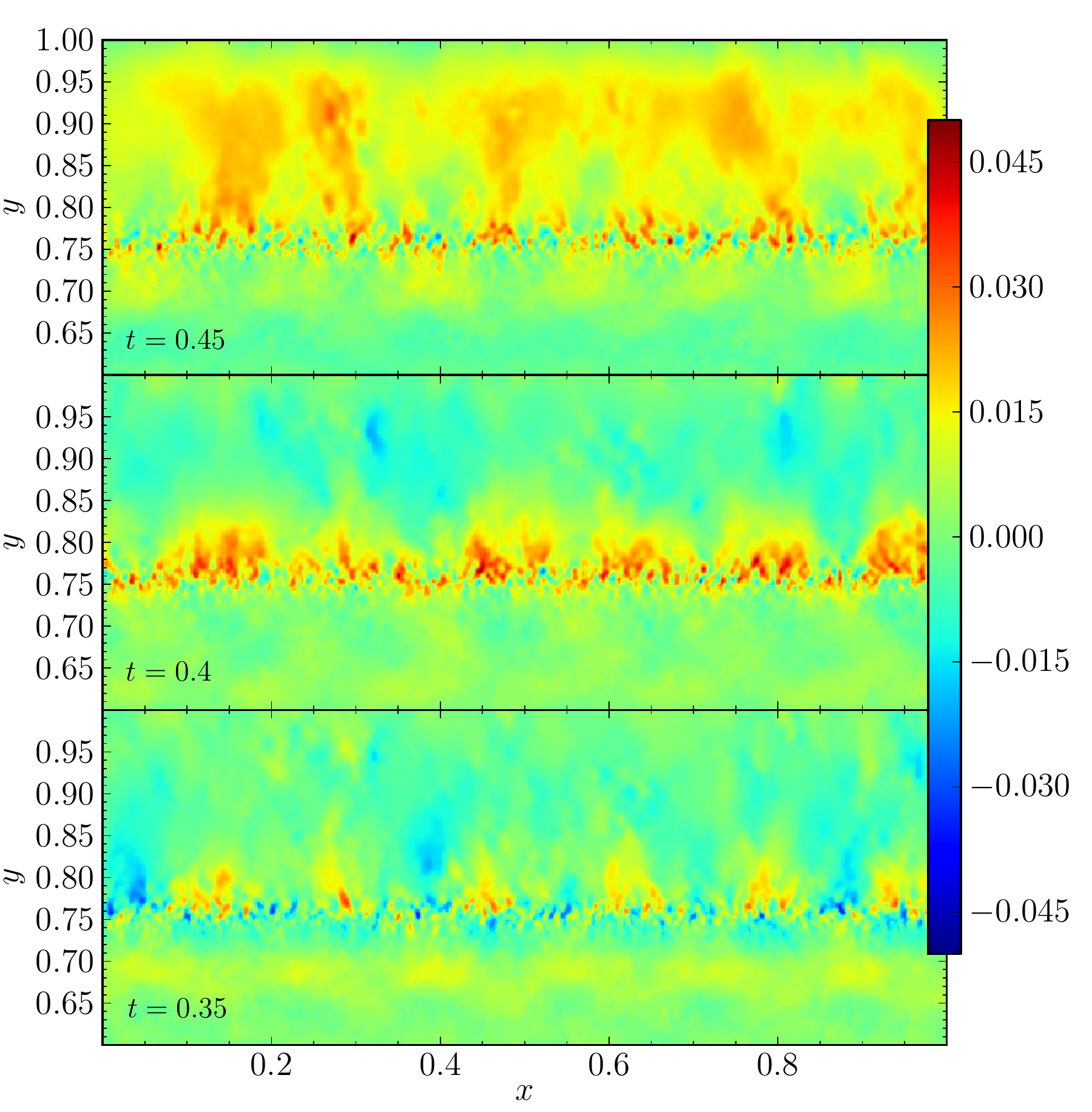}
\caption{$M=0.2$ simulation starting from a LIQ1 IC which was relaxed
  during a time of $0.5$. From bottom to top: $t=0.35, 0.4, 0.45$
  (after the end of relaxation). The same shock wave seen in
  Fig. \ref{fig:3406_vy} is present. However, its magnitude is greatly
  reduced, due to the relaxing of the IC. The initial KH seeds are
  also reduced in magnitude due to the increased particle disorder and
  general widening of the discontinuities.
\label{fig:smallRelaxVy}}
\end{figure}

\begin{figure}
\includegraphics[width=\hsize]{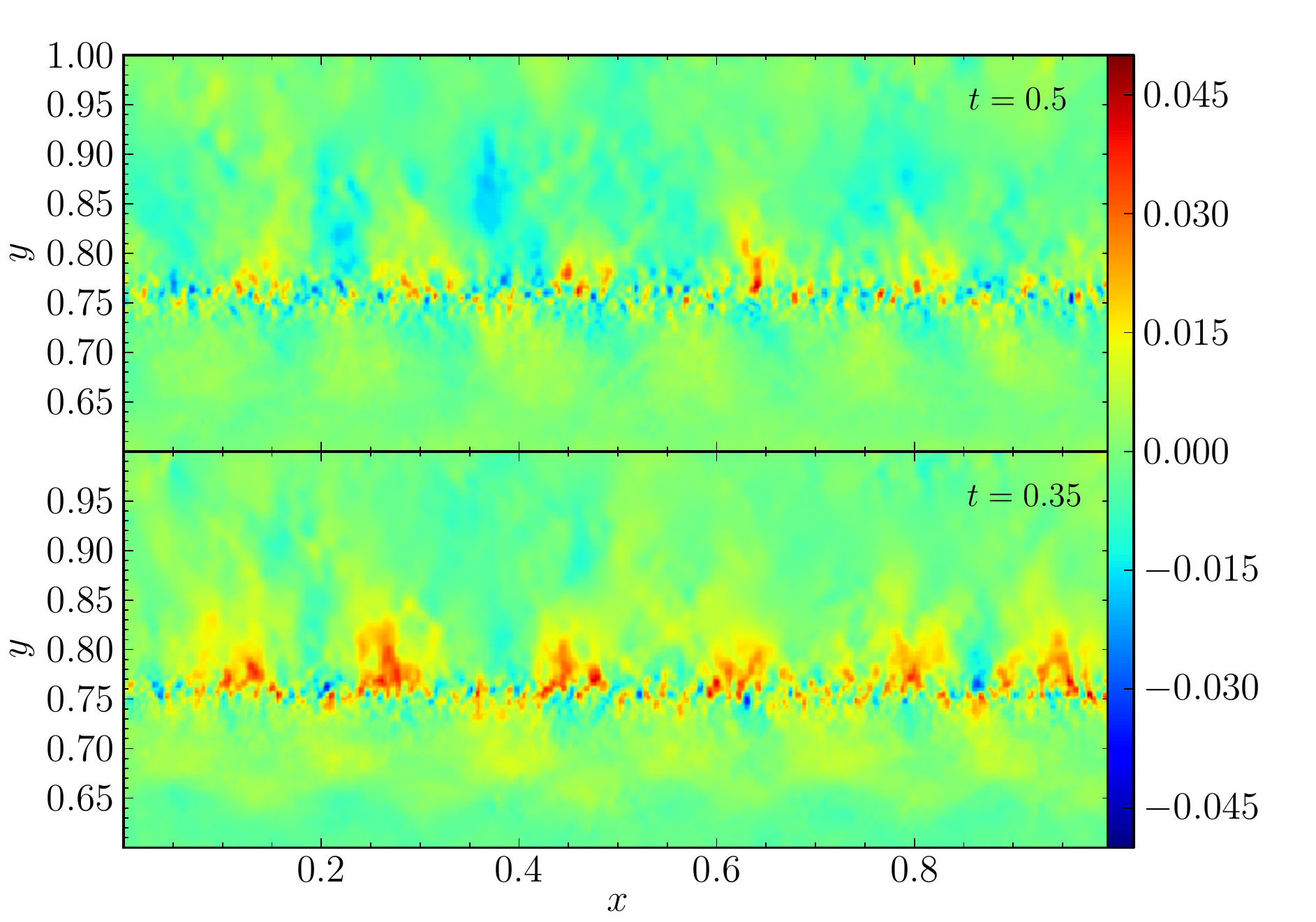}
\caption{$M=0.2$ simulation starting from a LIQ1 IC which was relaxed
  during a time of $1$. {\em Bottom panel}: $t=0.35$ (after the end of
  relaxation). {\em Top panel}: $t=0.5$ (after relaxation). No shock
  is seen, the initial particle disorder and widening of the
  discontinuities themselves remove the growing KH
  instabilities. \label{fig:fullRelaxVy}}
\end{figure}

We have relaxed run LIQ1. Two different runs were started from this
relaxing run, the first starting from the relaxing run snapshot at
$t=0.5$, the second from the $t=1.0$ snapshot. None of these
simulations show any formation of KH rolls. Rendered views of $v_y$
are shown in respectively Figs. \ref{fig:smallRelaxVy} and
\ref{fig:fullRelaxVy}. In Fig \ref{fig:smallRelaxVy}. we can see that
the magnitude of the shock (bottom panel, around $y=0.7$) is greatly
reduced compared to the shock in Fig. \ref{fig:3406_vy}. The emerging
instabilities are also reduced in magnitude when compared to those in
Fig. \ref{fig:3406_vy}, due to the increased particle disorder and
widening of the discontinuities. The small shock is sufficient to
remove these smaller emerging instabilities. In
Fig. \ref{fig:fullRelaxVy} we can see that the magnitude of the KH
seeds is very small. They do not develop and slowly fade away. This is
the result of the long relaxing time which greatly diffuses the energy
discontinuity and introduces too much particle disorder.

\subsection{Artificial Conductivity?}
\begin{figure*}
\includegraphics[width=\hsize]{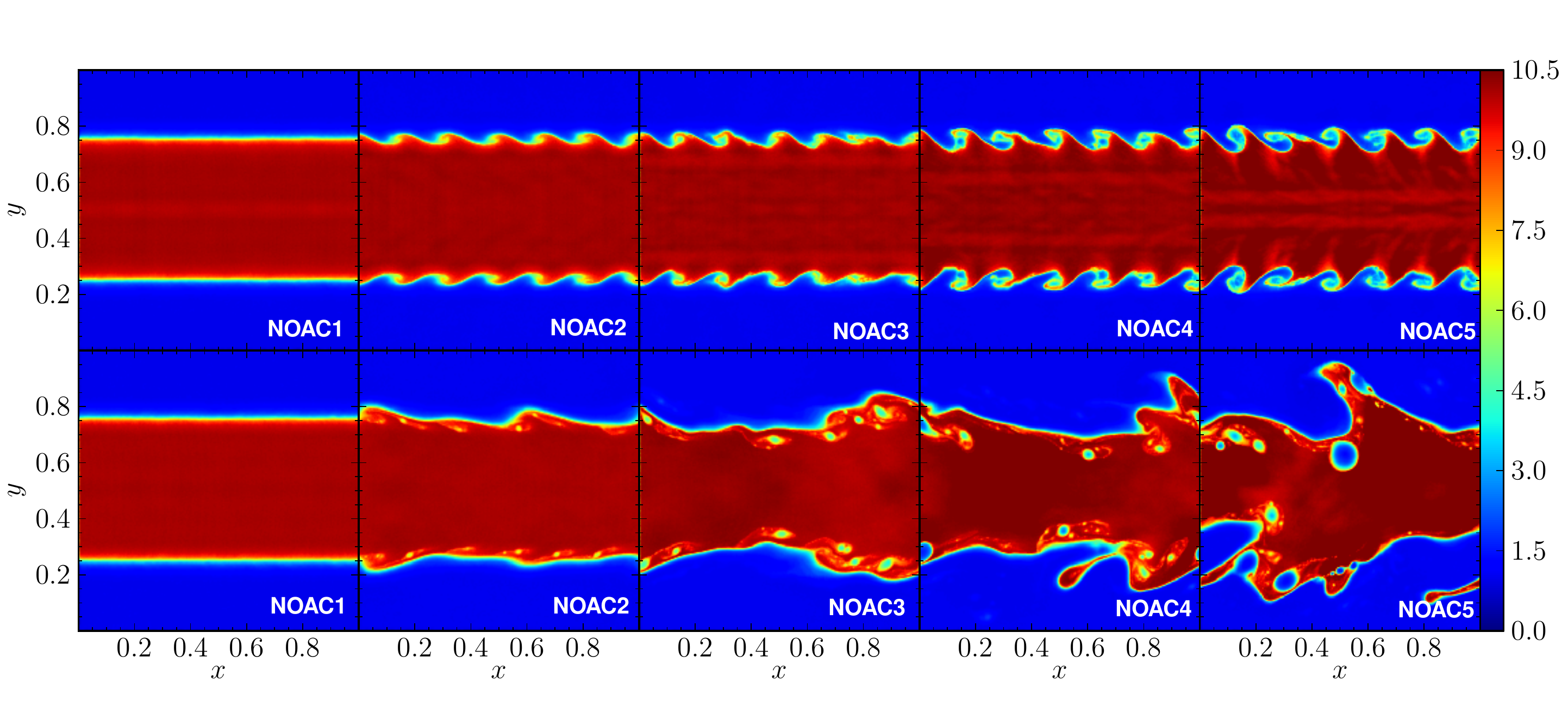}
\caption{Shearing layers simulations without artificial conductivity
  (column-smoothed density). {\em Left to right}: Simulations
  NOAC1-NOAC5. {\em Top row}: simulations at their respective
  $\tau\rs{KH}$. {\em Bottom row}: simulations at $t=2$.
\label{fig:density_noac}}
\end{figure*}

We have tested the need for artificial conductivity to be included in
shearing layers simulations using the LIQ kernel and with an initially
smoothed setup. To do this we repeated the LIQ1-5 simulations, with AC
switched off. Results are shown in Fig. \ref{fig:density_noac}. The
comparison with the bottom row of figure \ref{fig:sph_KH} learns us
several things: i) applying artificial conductivity does not lead to
the appearance of KH rolls. It is possible that with initial
conditions which are not smooth and when using the CS kernel AC,
provides an amount of smoothing in time for KHs to develop where they
would not otherwise. ii) Artificial conductivity is needed to avoid an
``oily'' nature of the gas. iii) When comparing the long-term
evolution of the simulations (Figs. \ref{fig:sph_KH_2} and
\ref{fig:density_noac}), artificial conductivity in its default
implementation is both a blessing and a curse. The panels in the
bottom row of Fig. \ref{fig:density_noac} show small-scale structures:
``holes'', inclusions of low-density gas inside the high-density
layer, are present in the high-density medium. When comparing with the
grid-based simulations (Figs. \ref{fig:elke_kh_02} and
\ref{fig:elke_kh_06}) we see that these are also present there. In the
simulation with AC applied these holes have entirely disappeared. It
is however obvious that the tendency for the layers to avoid mixing
without AC prevents realistic long-term behavior.
\subsubsection{Gap?}
\begin{figure}
\includegraphics[width=\hsize]{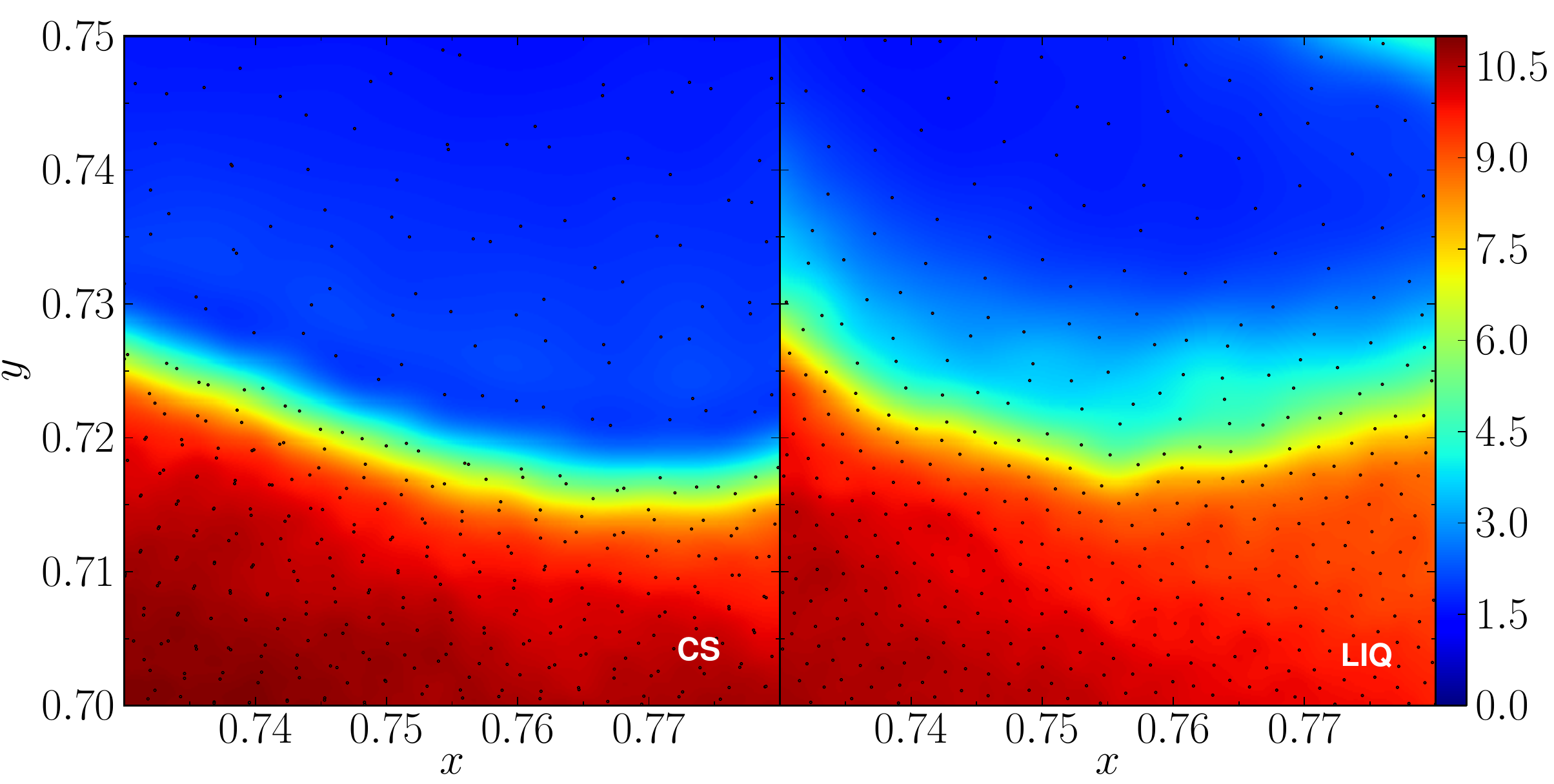}
\caption{Density of two shearing layers simulations with $M=1$, at
  $\tau\rs{KH}$, both without artificial conductivity. {\em Left
    panel}: Cubic spline kernel. {\em Right panel}: LIQ kernel. As
  previously shown the particle grouping of the CS kernel is abscent
  when using the LIQ kernel. Using the LIQ kernel also prevents the
  formation of a wide ``gap'' between the layers.
\label{fig:gap}}
\end{figure}

As recently highlighted by \citet{agertz2007}, SPH in its standard
form (without AC) suffers from the formation of a ``gap'', i.e. a
small void layer between two layers with different densities. In
Fig. \ref{fig:gap} we show plots of two simulations without artificial
conductivity, one with the CS kernel, the other with the LIQ
kernel. It is clear that the LIQ kernel prevents the formation of a
wide gap. As previously shown there still is a need for artificial
conductivity (see Fig. \ref{fig:density_noac}) because the layers are
still reluctant to mix. As we already achieved a great deal of
improvement by trying one new smoothing kernel, we do not think it
impossible that still better formulations are possible.
\subsubsection{Signal velocity}
\begin{figure}
\includegraphics[width=\hsize]{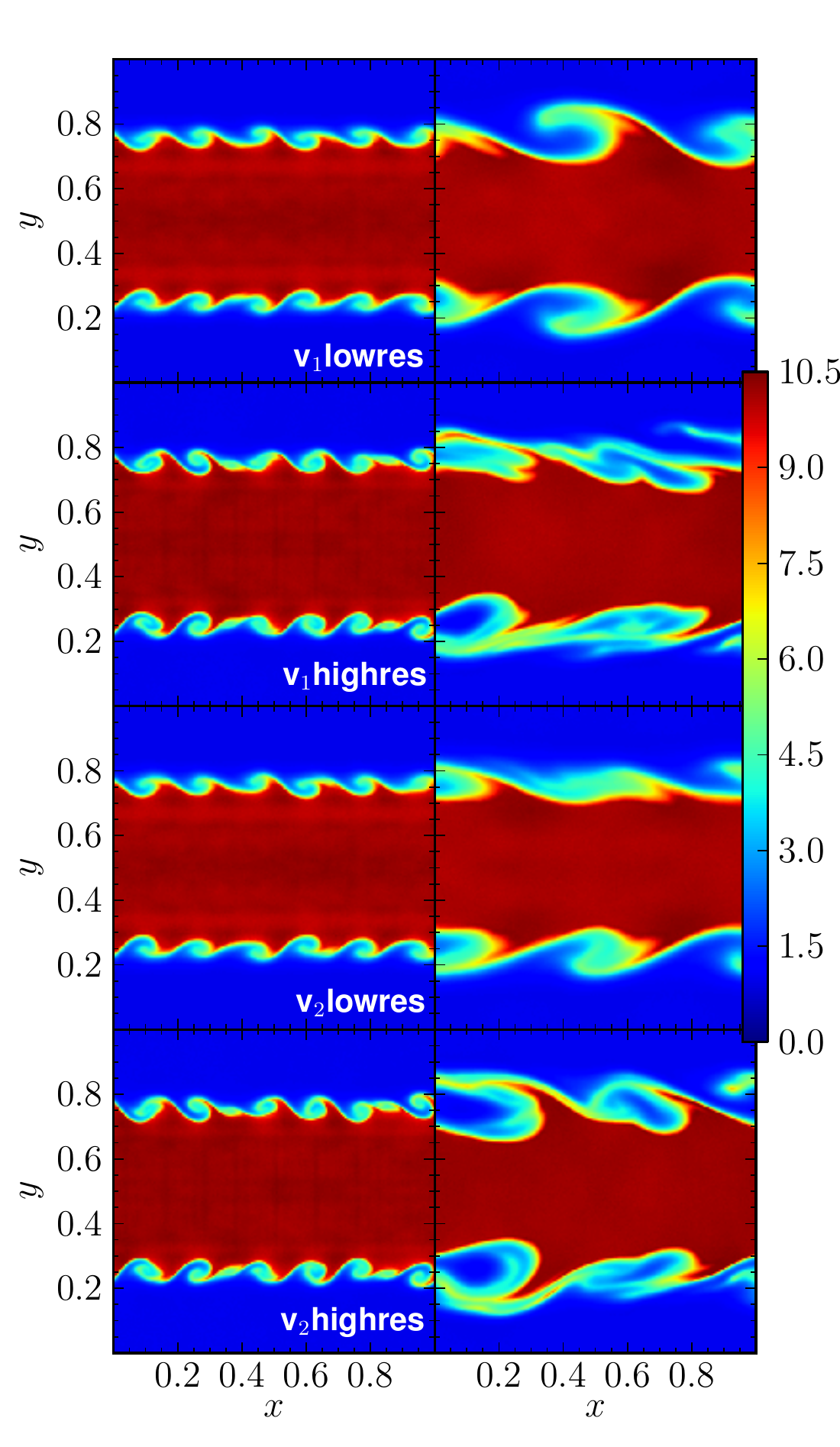}
\caption{Density plots of shearing layers simulations with LIQ3 setup
  (see tables \ref{tab:khsims} and \ref{tab:khsims2}, $M=0.6$), using
  different signal velocities. {\em Left column}:
  $t=\tau\rs{KH}$. {\em Right column}: $t=2$. Top two rows (label
  $v_1$): use the signal velocity from eq. (\ref{eq:ownvsig}). Bottom
  two rows (label $v_2$): use the signal velocity from
  eq. (\ref{eq:ownvsig2}). {\em lowres}: 184\,100 particles. {\em
    highres}: 552\,100 particles.
\label{fig:modvsig}}
\end{figure}

\begin{figure*}
\includegraphics[width=\hsize]{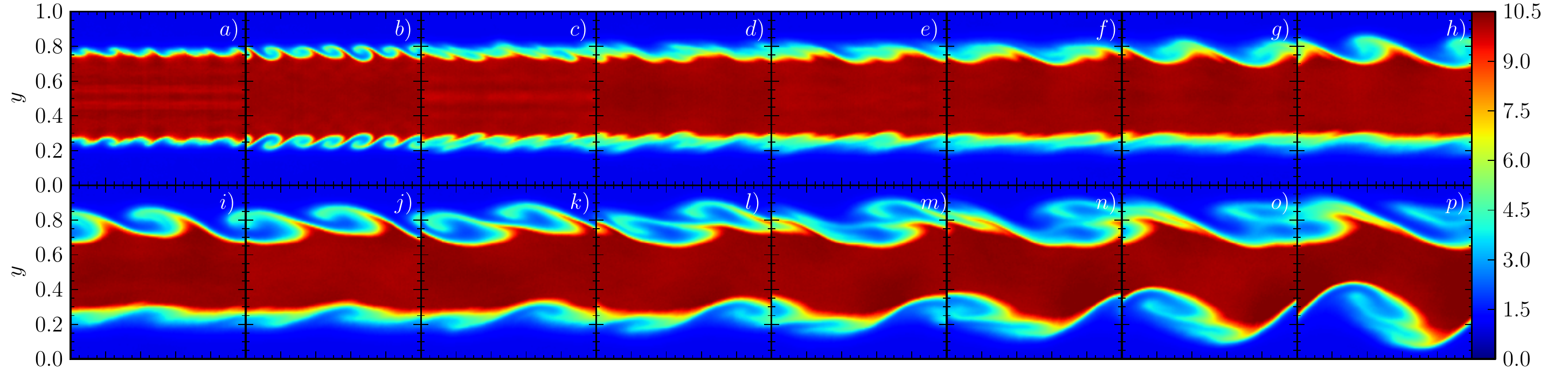}
\includegraphics[width=\hsize]{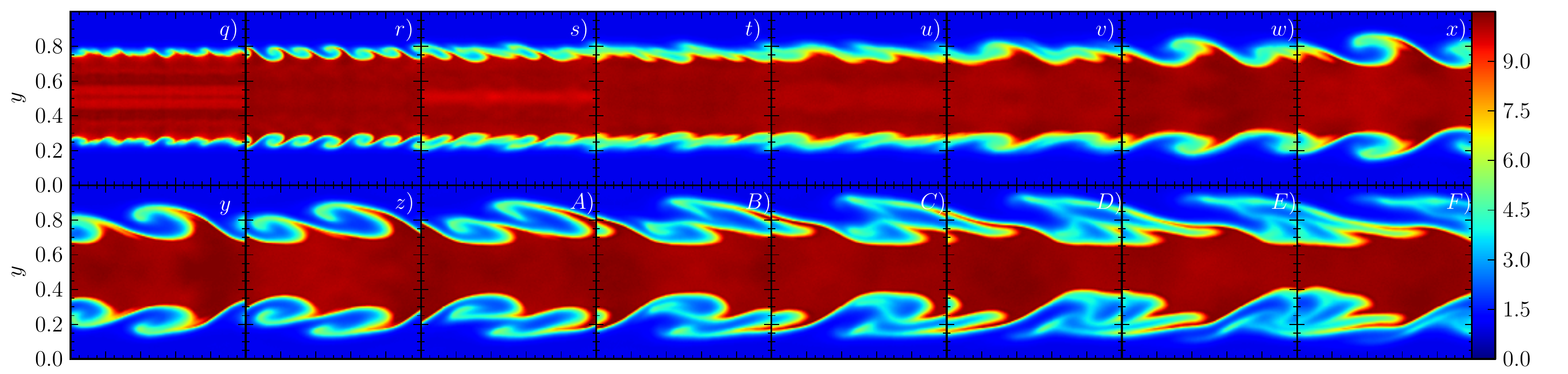}
\includegraphics[width=\hsize]{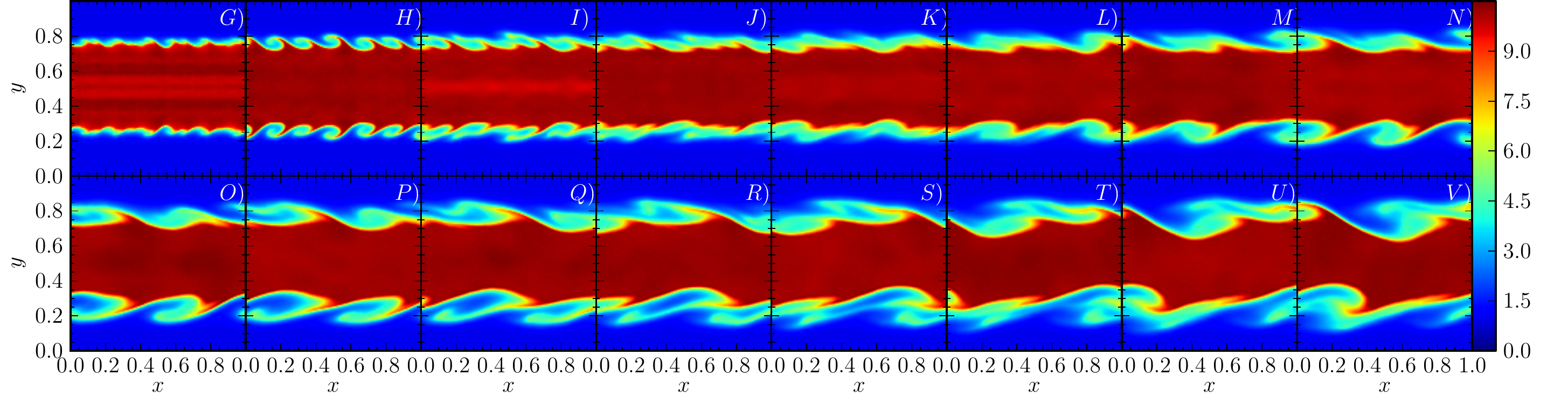}
\caption{16 snapshots of the density of 3 shearing layers
  simulations. Top rows (a-h, q-x, G-N): $t =
  0.25,0.5,0.75,1,1.25,1.5,1.75, 2$, bottom rows (i-p, y-F, O-V):
  $t=2.25,2.5,2.75, 3,3.25,3.5, 3.75,4$. {\em panels a-p}: simulation
  LIQ3. {\em panels q-F}: simulation LIQ3 with modified AC signal
  velocity $v\rs{sig,1}^u$ (see also Fig. \ref{fig:modvsig},
  v\rs{1}lowres). This simulation develops clearly resolved
  $\lambda=1/2$ KH rolls. {\em panels G-V}: simulation LIQ3 with
  modified AC signal velocity $v\rs{sig,2}^u$ (see also
  Fig. \ref{fig:modvsig}, v\rs{2}lowres). This develops less well
  defined $\lambda=1/2$ KH rolls, these do appear on both sides of the
  central stream, contrary to the LIQ3 simulation.
\label{fig:kh_series}}
\end{figure*}

In section \ref{subsec:theCode_AC} we briefly discussed the signal
velocity used when including AC in simulations. In
Fig. \ref{fig:modvsig} we show some results using the modified signal
velocities for simulations with $M=0.6$ and LIQ3 setup. They should thus
be compared to panel m of Fig. \ref{fig:sph_KH_2}. The top two rows show
simulations using $v\rs{sig,1}^u$ (eq. \ref{eq:ownvsig}), the bottom two
rows show results using $v\rs{sig,2}^u$ (eq. \ref{eq:ownvsig2}). For
both signal velocities, the results for $t=\tau\rs{KH}$ are in good
agreement with the grid simulation results as well as the other SPH
results. At $t=2$ the results differ from the grid results, with the
$\lambda=1/2$ clearly surfacing in the SPH simulations whereas this
instability takes longer to surface in the grid results. When comparing
with Fig. \ref{fig:sph_KH_2} it is obvious that for both signal
velocities the amount of energy diffusion applied is smaller. Indeed,
there is much more small-scale structure left in the $t=2$ panels in
Fig. \ref{fig:modvsig}.
We show a series of snapshots at different times in
Fig. \ref{fig:kh_series}, in order to have a clear view of which
instabilities are surfacing when. The top two rows (panels a-p) show the
evolution of the LIQ3 simulation. It goes from $\lambda=1/6$ to
$\lambda=1/2$ rolls, although the latter only appear on one side of the
high-density layer. Panels q-F show a LIQ3 simulation with
$v\rs{sig,1}^u$ AC signal velocity (eq. (\ref{eq:ownvsig})). Changing
the sign of the signal velocity is a clear improvement in this case,
with the KH rolls being even better resolved than in the LIQ3 simulation
and with the rolls appearing on both sides of the central flow (see the
grid simulation results in Fig. \ref{fig:elke_kh_06}). Panels G-V show a
LIQ3 simulation with $v\rs{sig,2}^u$ signal velocity
(eq. (\ref{eq:ownvsig2})). The KH rolls at $t=2$ are less well defined
than in the LIQ3 simulation, they do appear on both sides of the
flow. The result using $v\rs{sig,1}^u$ thus compares favorably to the
result using $v\rs{sig,2}^u$ and using $v\rs{sig,1}^u$ requires
virtually no extra computations, as opposed to using $v\rs{sig,2}^u$. It
seems that using $v\rs{sig,1}^u$ allows AC to act sufficiently strong to
prevent clear layer separation (``oiliness'' due to the LMI), but not
too strong so as not to lose too much resolution by diffusing too much
energy. Note that Fig. \ref{fig:modvsig} shows that the actual results
can change drastically when using a different resolution.

%% file: gridcodesection.tex
\subsection{Grid Code}
\label{subsec:gridcode}
To get a reference to compare the SPH results with we have conducted a
series of Eulerian hydrodynamical simulations, using the FLASH code
(version 3.2, (\citet{Fryxell2000flash,Dubey2008}, see also
\texttt{http://flash.uchicago.edu})) . FLASH is an adaptive mesh
refinement hydrodynamics code. Its standard hydro-solver uses the
piecewise-parabolic method.

The KH simulation is run on a periodic 2D grid of size $1\times 1$. In
order to avoid confusion from numerical issues as much as possible, we
switched off the steepening algorithm for contact discontinuities and
use just one refinement level, resulting in an effective resolution of
$256^2$ = 65536 grid cells.  We note that differences to runs with 2
refinement levels and contact steepening appear only after a few KH
timescales, but are evident in the late stages of the higher Mach number
runs presented here. The initial conditions are set as in the SPH
simulations, without taking special care to smooth gradients, but
keeping the idealised sharp discontinuity in density and velocity.
Results are shown in Figs. 6 and 7 for two simulations with respective
Mach numbers of 0.2 and 0.6. Those KH rolls that appear do so at times
consistent with their respective $\tau\rs{KH}$.

Results are shown in Figs. \ref{fig:elke_kh_02} and
\ref{fig:elke_kh_06} for two simulations with respective Mach numbers
of 0.2 and 0.6. Those KH rolls that appear do so at times consistent
with their respective $\tau_{\mathrm{KH}}$ .

%% file: discussion.tex
\section{Discussion}
\label{sec:discussion}
Using a large suite of simulations we have shown that the performance of
SPH on the shearing layers test can be improved drastically, if i) the
density gradient in the initial conditions is smoothed and ii) a
smoothing kernel that does not cause particle clumping is used. Two
effects are in play that can cause things to go haywire: shocks
travelling through the simulation box and particle clumping, or more
general, particle disorder. The effect of SPH particle disorder was
already discovered by \citet{okamoto2003}, who set up a simulation with
a small, hot shearing flow layer inside a cold medium. They found that
noisiness in the SPH smoothing of variables gives rise to small-scale
pressure gradients which significantly decelerated the shearing flow. As
using the Cubic Spline smoothing kernel causes particles to clump
together in groups of 2, as can be seen in the left panel of
Fig. \ref{fig:gap}, we can expect deceleration of SPH particles located
in a small layer around the contact interface. This layer then acts as a
lubricant between the two shearing layers, removing direct contact of
the latter two and hence preventing KH instabilities from forming or
growing. The LIQ kernel on the other hand gives rise to a much more
homogeneous particle distribution (right panel of
Fig. \ref{fig:gap}). Employing a suitble smoothing kernel with non-zero
central first derivative (we say suitable because although the first
derivative being non-zero is necessary to avoid clumping, it is not
sufficient: a LIQ kernel with connection point $x\rs{s}=0.5$ suffers
from even heavier clumping than the CS kernel: particles tend to clump
together in groups of 6) gives rise to a much more homogeneous particle
distribution. This has a major impact on shearing layers simulations,
allowing KHIs to form much easier in general. The effect of particle
disorder has been highlighted in the literature on various
occasions. \citet{read2010} show that particle clumping gives a large E0
error in the SPH momentum equation, which in turn prevents mixing in
SPH. \citet{morris1996} also found that that SPH does not converge at
flow boundaries because of the E0 error.

The shock problem is not so easily tackled. The shocks are triggered by
the local mixing instability, which occurs because energy (entropy) is
not smoothed in SPH. Several authors have identified this problem
\citep[e.g.][]{cummins1999, tartakovsky2005, agertz2007, price2008,
read2010} and various solutions have been suggested. We examined the
results using the artificial conductivity solution
\citep{price2008}. The magnitude of the shock-waves can be reduced by
applying an initial density smoothing, which reduces the magnitude of
any shock waves that might develop, leading to a significant increase in
the simulation results. To get rid of the remaining shocks one can relax
a simulation before applying velocities to it. This does not help in
this case however because relaxing not only significantly widens the
density and energy discontinuities, it also removes a great deal of the
symmetry that was initially present in the problem, i.e. it introduces a
lot of particle disorder. Indeed, a simulation started from a grid has
perfect initial particle symmetry and as sharp a discontinuity as one
desires. When relaxing the initial conditions on the other hand,
particle positions are shifted to reach an equilibrium. The smoothing
kernel plays an important part in this because through small-scale
variations in forces it will determine the final configuration of the
particles. This can readily be seen in
Fig. \ref{fig:shockTubeRender}. Even when using the LIQ kernel there is
an increased amount of particle noise. When bulk shearing layers
velocities are then given to particles based on on which side of a line
they lay, and velocity perturbations applied to particles irrespective
of the underlying configuration of the particles, it is not hard to
imagine that the problem of the particle noise will quickly result in
momentum transfer at the contact layers, thus shutting down all
KH-related activities. Furthermore, we found that the onset of the
shock-waves, occuring because of the LMI triggered in the initial
particle configuration, is not removed by the artificial conductivity.

We found that adding artificial conductivity to the SPH scheme does not
have an impact on whether or not initial ($\lambda=1/6$) KHIs
surface. Although it has been reported and used as such in the
literature \citep[see e.g.][]{price2008, kawata2009} we show that
simulations that use a suitable smoothing kernel and a smoothed initial
density gradient are equally able to form these KHIs, irrespective of AC
being included or not. Even the formation of a visible ``gap''
\citep{agertz2007} is prevented by using a suitable smoothing kernel
(Fig. \ref{fig:gap}). Including AC is, with the smoothing kernels used
in this paper, still necessary to i) allow mixing to happen, avoiding
``oily'' features in the SPH gas phases (see
Fig. \ref{fig:density_noac}) and preventing the LMI from triggering
during the course of the simulation and ii) get the $\lambda=1/2$ KH
rolls later on. Here, point ii) might very well be a consequence of
point i). We find that it is easy to actually lose substructure that was
initially resolved in SPH because of superfluous energy diffusion (see
Figs. \ref{fig:sph_KH_2}, \ref{fig:density_noac} and
\ref{fig:modvsig}). When using SPH to solve very sensitive
hydrodynamical problems (like the current shearing layers problem) one
should therefore take good care when including the artificial
conductivity and in selecting an appropriate signal velocity. We
presented two new signal velocities to that effect: $v\rs{sig,1}^u$
(eq. (\ref{eq:ownvsig})) and $v\rs{sig,2}^u$
(eq. (\ref{eq:ownvsig2})). Shearing layers results indicate that both
these signal velocities lead to less energy diffusion in
simulations. $v\rs{sig,1}^u$ emerges as the best choice, giving
better results in terms of KH rolls and requiring no significant extra
computations.

The combined effect of both the shock waves and the particle disorder
becomes less and less important as the time-scale of the problem itself
(in this case: $\tau\rs{KH}$) decreases. At he resolution of current
galaxy formation simulations mixing is probably not important. However,
mixing could become crucial for next-generation simulations. Also, for
the standard astrophysical application the actual choice of AC sigal
velocity will be less of a concern, if a concern at all, as is
demonstrated by the host of successful problems tackled by the SPH codes
of \citet{rosswog2007} and \citet{kawata2009}. Note that a suitable
signal velocity for simulations including gravity has not as of yet been
presented.

%% file: appendix.tex
\section{LIQ kernel coefficients}
\label{sec:appendixLiq}
The expressions for the parameters of the LIQ kernel
(eq. (\ref{eq:liqKernel})), obtained by solving the set of equations
(\ref{eq:liq_firstEq}--\ref{eq:liq_lastEq}), are:
\begin{eqnarray}
  \alpha &=& \frac{1}{x\rs{s}^3 - 3x\rs{s}^2 + 3x\rs{s} - 1} \\
  A      &=& \frac{\alpha}{2} \\
  B      &=& - \alpha(1 + x\rs{s}) \\
  C      &=& 3 \alpha x\rs{s} \\
  D      &=& -\alpha(-1 + 3x\rs{s}) \\
  E      &=& \frac{\alpha(2 x\rs{s} - 1)}{2} \\
  F      &=& A x\rs{s}^4 + B x\rs{s}^3 + C x\rs{s}^2 + D x\rs{s} + E + x\rs{s},
\end{eqnarray}
with $x\rs{s}$ the LIQ kernel connection point. From this it is
straightforward to calculate the norm $N$ by integrating over the
volume:
\begin{equation}
  N = \left[ \int_0^{x\rs{s}} W_r(u)\mathrm{d}u + \int_{x\rs{s}}^1 W_r(u) \mathrm{d}u \right]^{-1}.
\end{equation}
Calculating $N$ is then straightforward. We give the expressions here
for completeness. In two dimensions:
\begin{align}
  \int_0^{x\rs{s}} W_r(u) \mathrm{d}u &= 2\pi\left(\frac{1}{2}Fx\rs{s}^2 - \frac{1}{3}x\rs{s}^3\right) \\
  \int_{x\rs{s}}^1 W_r(u) \mathrm{d}u &=  2\pi\Big( \frac{A}{6}x\rs{s}^6 + \frac{B}{5}x\rs{s}^5 
     +\frac{C}{4}x\rs{s}^4 \\ 
     &\qquad\qquad\qquad + \frac{D}{3}x\rs{s}^3 +\frac{E}{2}x\rs{s}^2\Big)\Big|_{x\rs{s}}^1.
\end{align}
In three dimensions:
\begin{align}
  \int_0^{x\rs{s}} W_r(u) \mathrm{d}u &=  4\pi\left(\frac{1}{3}Fx\rs{s}^3 - \frac{1}{4}x\rs{s}^4\right) \\
  \int_{x\rs{s}}^1 W_r(u) \mathrm{d}u &=  4\pi\Big( \frac{A}{7}x\rs{s}^7 + \frac{B}{6}x\rs{s}^6
     +\frac{C}{5}x\rs{s}^5 \\ 
     &\qquad\qquad\qquad + \frac{D}{4}x\rs{s}^4 +\frac{E}{3}x\rs{s}^3\Big)\Big|_{x\rs{s}}^1.
\end{align}